\documentclass[journal,10pt]{IEEEtran}
\usepackage{graphicx, epstopdf}
\usepackage{amssymb}
\usepackage{amsmath}
\usepackage{amsthm}
\usepackage{mathrsfs}
\usepackage{cite}
\usepackage{color}
\usepackage[acronym]{glossaries}
\usepackage[T1]{fontenc}
\usepackage[latin9]{inputenc}
\usepackage{array}
\usepackage{textcomp}
\usepackage{multirow}
\usepackage{soul}

\newtheorem{lem}{Lemma}

\newtheorem{remark}{Remark}

\newtheorem{proposition}{Proposition}

\makeatletter
\newcommand*\titleheader[1]{\gdef\@titleheader{#1}}
\AtBeginDocument{%
  \let\st@red@title\@title
  \def\@title{%
    \bgroup\normalfont\small\centering\@titleheader\par\egroup
    \vskip0.5em\st@red@title}
}

\title{Ultra-small Cell Networks with Collaborative RF and Lightwave Power Transfer}
\author{
        {
        Ha-Vu Tran, Georges Kaddoum, {\it Member, IEEE}, Panagiotis D. Diamantoulakis, {\it Member, IEEE}, Chadi Abou-Rjeily, {\it Senior Member, IEEE}, and George K. Karagiannidis, {\it Fellow, IEEE}
         \vspace{-10pt}
        }
\thanks{
Ha-Vu Tran and Georges Kaddoum are with University of Qu\'{e}bec, \'{E}TS engineering school, LACIME Laboratory, Montreal, Canada  (e-mails: ha-vu.tran.1@ens.etsmtl.ca, georges.kaddoum@etsmtl.ca).

P. D. Diamantoulakis and G. K. Karagiannidis are with the Department of Electrical and Computer Engineering, Aristotle University of Thessaloniki, Thessaloniki, Greece (e-mails: \{padiaman,geokarag\}@auth.gr).

Chadi Abou-Rjeily is with the Department of Electrical and Computer Engineering of the Lebanese American University (e-mail: chadi.abourjeily@lau.edu.lb).

}
 }

\begin{document}
\makeatother

\titleheader{This is the authors'version of the paper that has been accepted for publication in IEEE Transactions on Communications.}
 \maketitle

\begin{abstract}
This paper investigates a hybrid radio frequency (RF)/visible light communication (VLC) ultra-small cell network consisting of multiple optical angle-diversity transmitters, one multi-antenna RF access point (AP), and multiple terminal devices. In the network, the optical transmitters play the primary role and are responsible for delivering information and power over the visible light, while the RF AP acts as a complementary power transfer system. Thus, we propose a novel collaborative RF and lightwave resource allocation scheme for hybrid RF/VLC ultra-small cell networks. The proposed scheme aims to maximize the communication quality-of-service provided by the VLC under a constraint of total RF and light energy harvesting performance, while keeping illumination constant and ensuring health safety. This scheme leads to the formulation of two optimization problems that correspond to the resource allocation at the optical transmitters and the RF AP. Both problems are optimally solved by appropriate algorithms. Moreover, we propose a closed-form suboptimal solution with high accuracy to tackle the optical transmitters' resource allocation problem, as well as an efficient semi-decentralized method. Finally, simulation results illustrate the achievable performance of the investigated system and the effectiveness of the proposed solutions.
\end{abstract}


\section{Introduction}
To cope with the exponentially growing demand for data traffic, the design of future cellular networks, i.e., the fifth generation (5G) networks, is tending to a new form, embracing a large-scale deployment of small cells \cite{Gupta2015,Ekram2015}. In this context, a potential paradigm shift of 5G networks appeared, so-called {optical attocell networks}, developed on the platform of visible light communication (VLC), lifting the small cell concept to a new level of ultra-small cells \cite{HaraldHaas2017, ZheChen2018,ChengChen2016}.  In particular, the advantages of this optical network are related to the fact that the visible light spectrum is 1000-fold wider than the entire radio frequency (RF) spectrum. Moreover, artificial light sources, i.e., light emitting diode (LED) bulbs, can be densely deployed in indoor areas, such as houses, offices, etc., with a conveniently lower cost than their RF counterparts, i.e., WiFi access points \cite{wifilifi2016,Pathak2015}.

Furthermore, while moving toward future networks, another main challenge is prolonging the network lifetime. More specifically, the lifetime of mobile devices is expected to be 10-fold longer, while the overall network power consumption should not exceed 10 percent of the current usage \cite{Ekram2015,DusitNiyato}. To cope with these expectations, energy harvesting (EH), wireless power transfer (WPT), and simultaneous wireless information and power transfer (SWIPT) can be considered as primary solutions \cite{Ekram2015,DusitNiyato,BrunoClerckx, Lu2015, Tran2017}. Particularly, SWIPT, when properly optimized, can result in significant gains in terms of spectral efficiency, time delay, energy consumption, and interference management, by superposing information and power transfer \cite{Krikidis2011,Timotheou2014}.

While RF WPT is a well explored technology, recently, along with the development of VLC \cite{Kavehrad2010, Mossaad2015, Arnon2015,Ma2015,Wang2015}, optical WPT and EH from artificial light have gained increasing attention from the research community \cite{Fakidis,Carvalho,Nasiri,Mathews,TamerRakia,HinaTabassum2018,Basnayaka2017,GaofengPan,Diamantoulakis2018}. It has been shown that wireless users in VLC EH networks can benefit from the illumination, data communication and light EH simultaneously. In the literature, there are some fundamental works providing measurement results regarding optical wireless power transfer \cite{Fakidis} as well as EH from indoor artificial light \cite{Carvalho,Nasiri,Mathews}. Further, the work in \cite{TamerRakia} has considered a dual-hop hybrid VLC/RF communication system; the relay harvests energy over the VLC first-hop link by taking the DC component out of the received signal and uses this energy for retransmitting the signal over the RF second-hop link. Some scenarios of hybrid VLC/RF networks with multiple RF access points (APs) and multiple optical transmiters have been studied and analyzed in \cite{HinaTabassum2018,Basnayaka2017}.  In \cite{GaofengPan}, the authors have proposed a secure scheme in which the harvested energy from indoor artificial light via downlinks is used for uplink communication. Moreover, the framework of simultaneous lightwave information and power transfer was introduced for the first time in \cite{Diamantoulakis2018}. Particularly, according to \cite{GaofengPan,Diamantoulakis2018}, the light energy harvested in VLC systems is around 1.5 mW which can provide enough operating energy for low-power devices, such as IoT sensors \cite{TomTorfs2013}. However, it might only supply 11\% of the energy needed to smartphones. Precisely, a smartphone requires an EH rate of 13.69 mW\footnote {In the provision of continuous services, it is expected that smartphones can always be rechared with sufficient energy for the phone call service. According to \cite{Carroll2010}, a phone consumes 1054.3 mW averaged over 77 seconds for a call which corresponds that 13.69 mW of harvested power is needed.}. 
Indeed, the energy demand is more than the energy offered by conventional VLC systems. Nevertheless, it is noted that the indoor light provides {\it free} energy since no extra power is needed from the lighting system. Moreover, a transparent solar panel named as Wysips Reflect, recently developed by Sunpartner Technologies \cite{wysipscrystal}, can be integrated into phone screens. It promises to remarkably enhance the light EH performance.
Therefore, the previous works \cite{Fakidis,Carvalho,Nasiri,Mathews,TamerRakia,HinaTabassum2018,Basnayaka2017,GaofengPan,Diamantoulakis2018,wysipscrystal} have shown the promising performance of using visible light for communications and power transfer.

Nevertheless, in VLC systems, there are three major issues of EH from indoor light. First, the amount of indoor light energy is limited. This is due to the fact that the intensity of LED light is lower than the one of solar light, even in well-illuminated areas, such as drafting tables or workshops \cite{Mathews}.  Second, during the non-working time at night, LED bulbs in office buildings are often turned off or dimmed to save energy. Thus, the EH performance is significantly decreased.  Third, the illumination in indoor living environments should respect eye safety standards. Therefore, LED light bulbs might not change flexibly their emitting directions or optical beamforming to maximize the EH performance. Different from visible light, RF is more flexible, and its well-known application, RF WPT, has been well studied in the literature \cite{DusitNiyato,Liu2014,HJu2014,Vuwcnc,SWIPTKaddoum,VuTVT2018,Clerckx2016,Boshkovska2015a,Boshkovska2017,RuihongJiang2017,KeXiong2017,MengLinKu2018, Tabassum2015,Tran2019wcl}. Thus, a combination between optical and RF WPTs can provide an efficient solution in both power and data transmisions, since it becomes evident that RF and lightwave wireless power transfer approaches are complementary rather than competing technologies. Indeed, the cooperation between the two technologies seems to be a particularly promising direction that can fundamentally extend the efficacy of single-band WPT (i.e., either RF or lightwave), while respecting the power constraints per band for safety and consistent illumination reasons. 

Motivated by the aforementioned issues, in this paper, we consider a hybrid RF/VLC network consisting of multiple optical transmitters, one multi-antenna RF AP, and multiple terminal devices. Each optical transmitter, composed of multiple LED elements, can generate multiple narrow beams simultaneously, known as optical angle-diversity transmitters \cite{ZheChen2018}. This configuration in terms of spatial division can enhance the bandwidth resource and eliminate intercell interference in attocell networks \cite{ZheChen2018}. Further, each terminal device is equipped with one single antenna and one photodetector. We assume that the devices have a multi-homing capability that allows to aggregate resources received from the optical transmitters and the RF AP \cite{Kashef2016}. Due to the limitations of light EH, the network can fail to ensure the desirable light EH performance while guaranteeing information decoding (ID) performance. On this basis, this work aims at proposing a novel collaborative RF and lightwave resource allocation scheme to enhance the EH performance of the overall network.

In the proposed scheme, the optical transmitters play the primary role and are responsible for delivering both information and energy over only visible light while the RF AP is a helper which is liable for transferring wireless power using RF. In the system model, the communication quality of service (QoS) is maximized for all users, in terms of VLC signal-to-noise ratios (SNRs), under a constraint of the sum of light and RF EH performance at each user, while keeping illumination constant and ensuring health safety in the network area. In this concern, the RF AP is configured to minimize its power budget while contributing a certain amount of the RF EH performance to the overall network. Thus, we formulate two corresponding optimization problems associated with the group of the optical transmitters and the RF AP, whose solving process are tightly related to each other. Hence, we devise algorithms to handle and optimally solve the two problems. Moreover, to reduce the computational burden, we derive a closed-form suboptimal solution to the problem of the group of the optical transmitters. Finally, considering the case where each optical transmitter is capable of calculating the suboptimal solution using the low complex closed-form expression, we further propose a semi-decentralized method to further facilitate the solving process.

In summary, the main contributions of our work can be summarized as follows:
\begin{itemize}
\item Proposing a novel collaborative RF and lightwave resource allocation scheme for hybrid RF/VLC ultra-small cell networks with simultaneous power and data transmission.
\item Devising algorithms to solve the optimization problems of the optical transmitters and the RF AP optimally.
\item Providing a closed-form suboptimal solution with high accuracy and low complexity to the problems of the optical transmitters. 
\item Developing a semi-decentralized method to tackle the problems of the network.
\end{itemize}

The remainder of this paper is organized as follows: 
In Section II, the system model is presented. 
The proposed collaborative resource allocation scheme and its formulation are shown in Section III.
The optimal and the suboptimal solutions are given in Section IV. In Section V, numerical results are provided and discussed. Finally, the conclusion is put forward in Section VI.

\section{System Model}
\subsection{Optical Angle-diversity Transmitters with Color Allocation}
In this work, we consider the optical transmitters composed of multiple LED elements which can generate multiple narrow beams simultaneously as in \cite{ZheChen2018}. Each LED element points in a distinct direction. The angle diversity transmitter, whose layout is described in \cite{ZheChen2018}, covers the same area and provides the same white illumination as conventional single-element transmitters. However, since the beam of each LED element can interfere with the others, multi-color white LED lamps, e.g. RGB white LEDs, can be a promising candidate to overcome this issue. Specifically, one of the three colors is selected to convey communication data whereas the others are used to keep illumination constant. Since the three colors are orthogonal, the interference can be canceled by properly selecting a color for each LED element.
Further, the unselected colors operate at zero alternating current (AC) and non-zero direct current (DC) bias. Consequently, they have no impact on the information detection of neighboring receivers.

\begin{figure}[t]
\centering
{\includegraphics[width=0.45\textwidth]{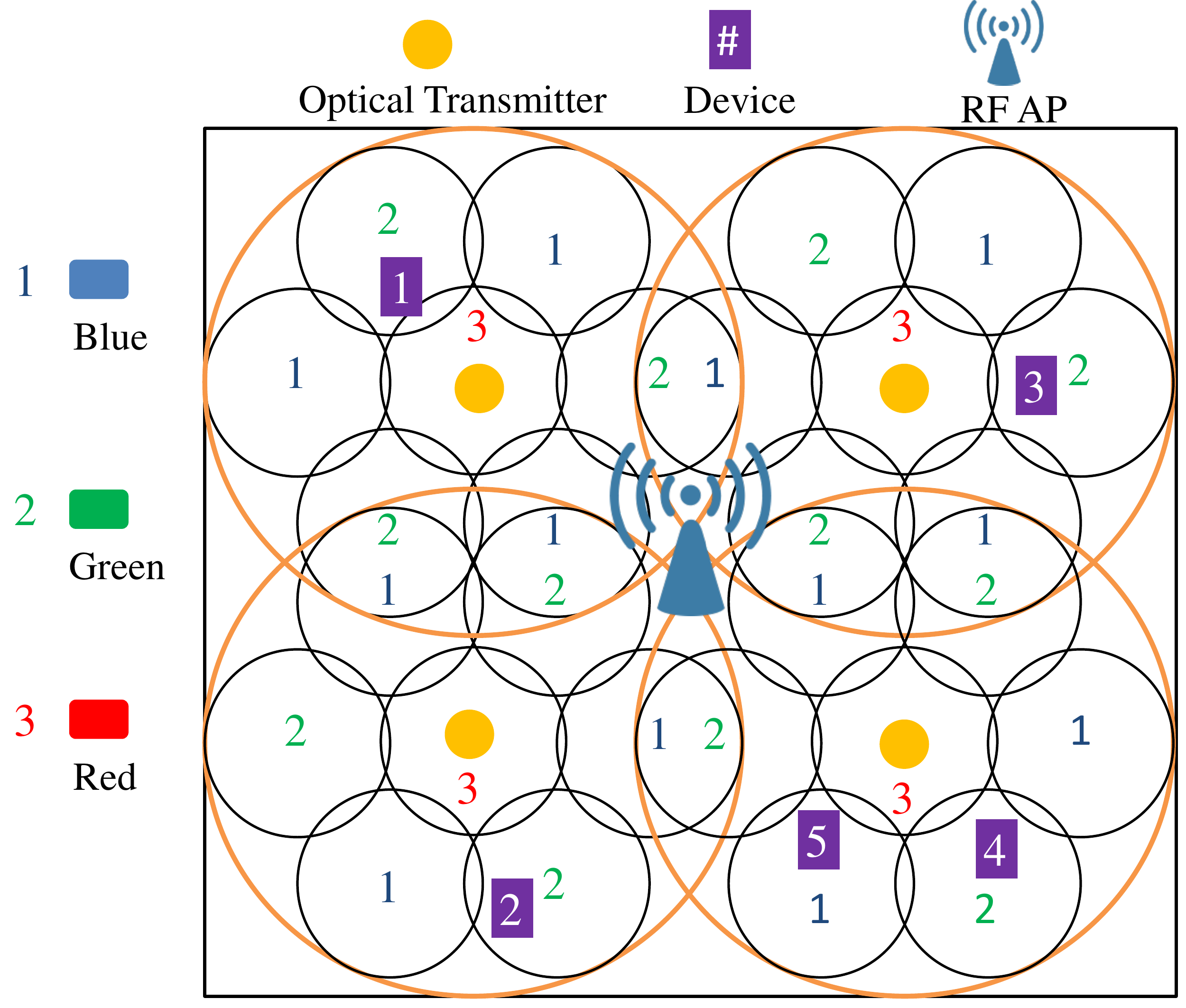}}
   \caption{
     An illustration of the network with color allocation.
    }
    \label{fig:SystemModel1}
\end{figure}

Given the use of the optical angle-diversity with color allocation, we consider a system model consisting of one RF AP with $M_T$ antennas, $\mathcal O$ optical transmitters with $M_I$-element angle-diversity, and $J$ terminal devices, as shown in Fig. \ref{fig:SystemModel1}. The optical transmitters are placed on the ceiling to provide illumination and communication at the same time. Each multi-homing terminal device is equipped with a single antenna and a solar panel. We assume that each device is served by an optical transmitter.

In Fig. \ref{fig:SystemModel1}, an illustration of the network with 7-element angle diversity optical transmitters is shown where the information-carrying colors are interleavedly allocated to circular areas to eliminate interference. Further, to avoid inter-transmitter interference, two vertically consecutive transmitters are adjusted to guarantee a shift of $\frac{360}{M_I}$ degrees between them. Particularly, it is worth noting that the light intensity or the radiance angle at half-power in the covered area is the same regardless of the values of ${M_I}$ \cite{ZheChen2018}. Under this condition, increasing the number of elements ${M_E}$ improves the spectral efficiency and increases the number of users able to be simultaneously served \cite{ZheChen2018}, however, it does not impact the lightwave WPT performance.

\begin{figure}[t]
\centering
{\includegraphics[width=0.5\textwidth]{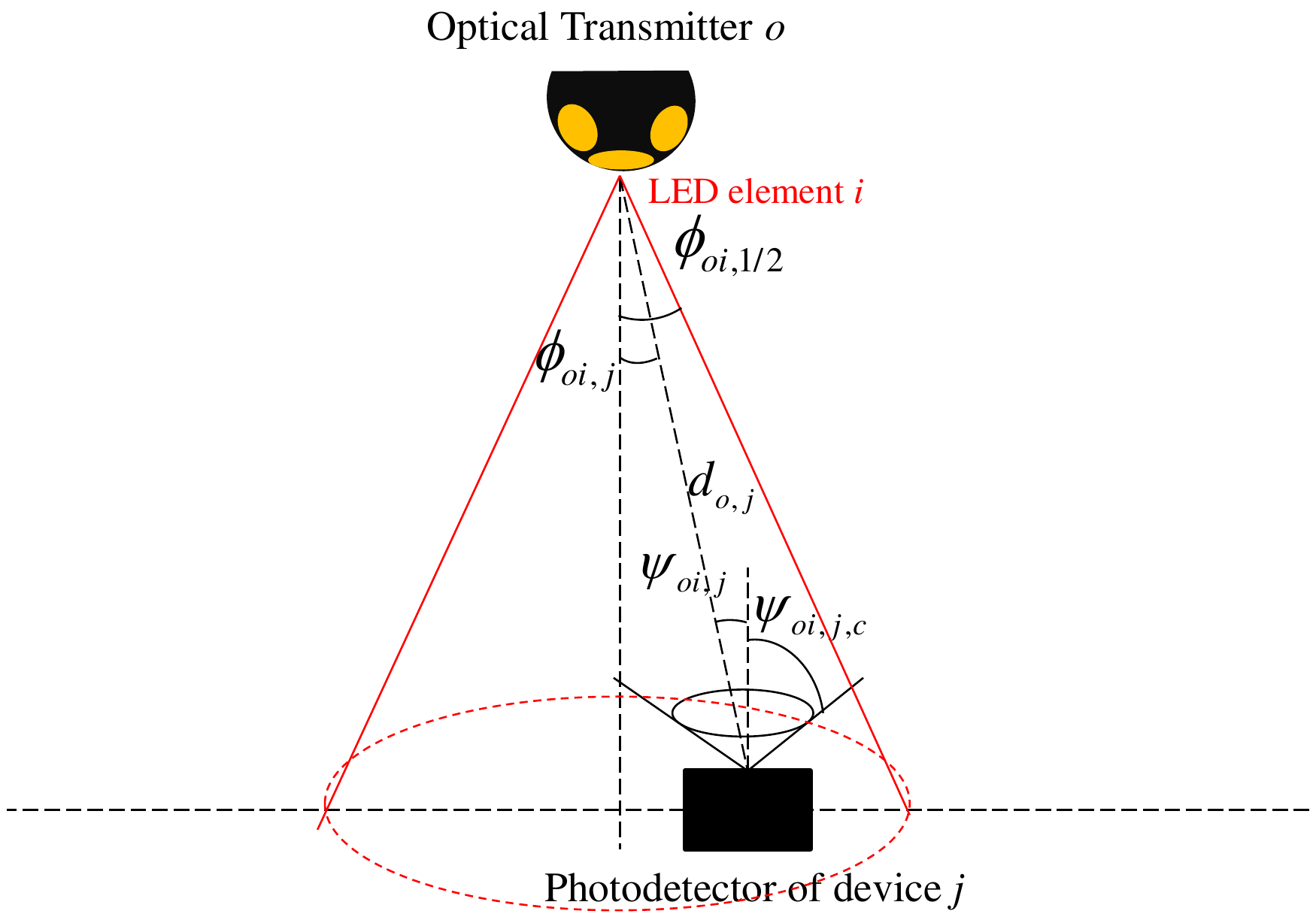}}
   \caption{
      Visible light communication channel.
    }
    \label{fig:LOS}
\end{figure}

\subsection{VLC Channel Model}
The total received optical power at a receiver can be composed of line-of-sight (LOS) and non-line-of-sight (NLOS) components. However, the contribution of the second component is much smaller than that of the first one \cite{GaofengPan,Diamantoulakis2018}. In this work, for simplicity, we consider only the VLC channels with the LOS component. Thus, the VLC channel between LED element $i$ $(1 \le i \le M_I)$ of optical transmitter $o$ $(1 \le o \le \mathcal O)$ and the photodetector of device $j$ $(1 \le j \le J)$, denoted by $h_{oi,j}$, can be given as \cite{GaofengPan,Diamantoulakis2018}
\begin{align}
h_{oi,j} &= \dfrac{Ar_j (m_{i} + 1)}{2\pi d_{oi,j}^2} \text{cos}^{m_{oi}}(\phi_{oi,j}) T_s(\psi_{oi,j} ) \nonumber \\
&\times g(\psi_{oi,j} ) \text{cos}(\psi_{oi,j} ), 
\end{align}
where $Ar_j$ is the active area of the photodetector, $d_{oi,j}$ is the distance between the LED element to the photodetector of device $j$, $\phi_{oi,j}$ is the irradiation angle, $\psi_{oi,j}$ is the incident angle of light radiation, $T_s(\psi_{oi,j} )$ is the optical band-pass filter gain of transmission, $m_{oi}$ is the Lambert's mode number, and $g(\psi_{oi,j} )$ is the optical concentrator gain. The different parameters of the VLC channel are shown in Fig. \ref{fig:LOS}. Furthermore, $\phi_{oi,1/2}$ is the LED semi-angle at half-power, and $\psi_{oi,j,c} \le \pi/2$ is the field of view (FOV). These two paramteters are used to compute $m_{oi}$ and $g(\psi_{oi,j} )$ \cite{GaofengPan,Diamantoulakis2018}.

\subsection{RF Channel Model}
In this paper, we consider the Rician channel model for RF links. The transmission channel between the RF AP and device $j$ is given by
\begin{align}
{\mathbf g}_j= \sqrt{ \frac{R}{1+R} } {\mathbf g}_{j,0}   + \sqrt{ \frac{1}{1+R} } {\mathbf g}_{j,1},   
\end{align}
where ${\mathbf g}_{j,0} \in \mathcal {CN}^{M_T \times 1}$ denotes the line of sight component, ${\mathbf g}_{j,1} \in \mathcal {CN}^{M_T \times 1}$ represents the non-line of sight fading component, and $R$ is the Rician factor.

\subsection{Signal Models}

\begin{figure}[!]
\centering
{\includegraphics[width=0.38\textwidth]{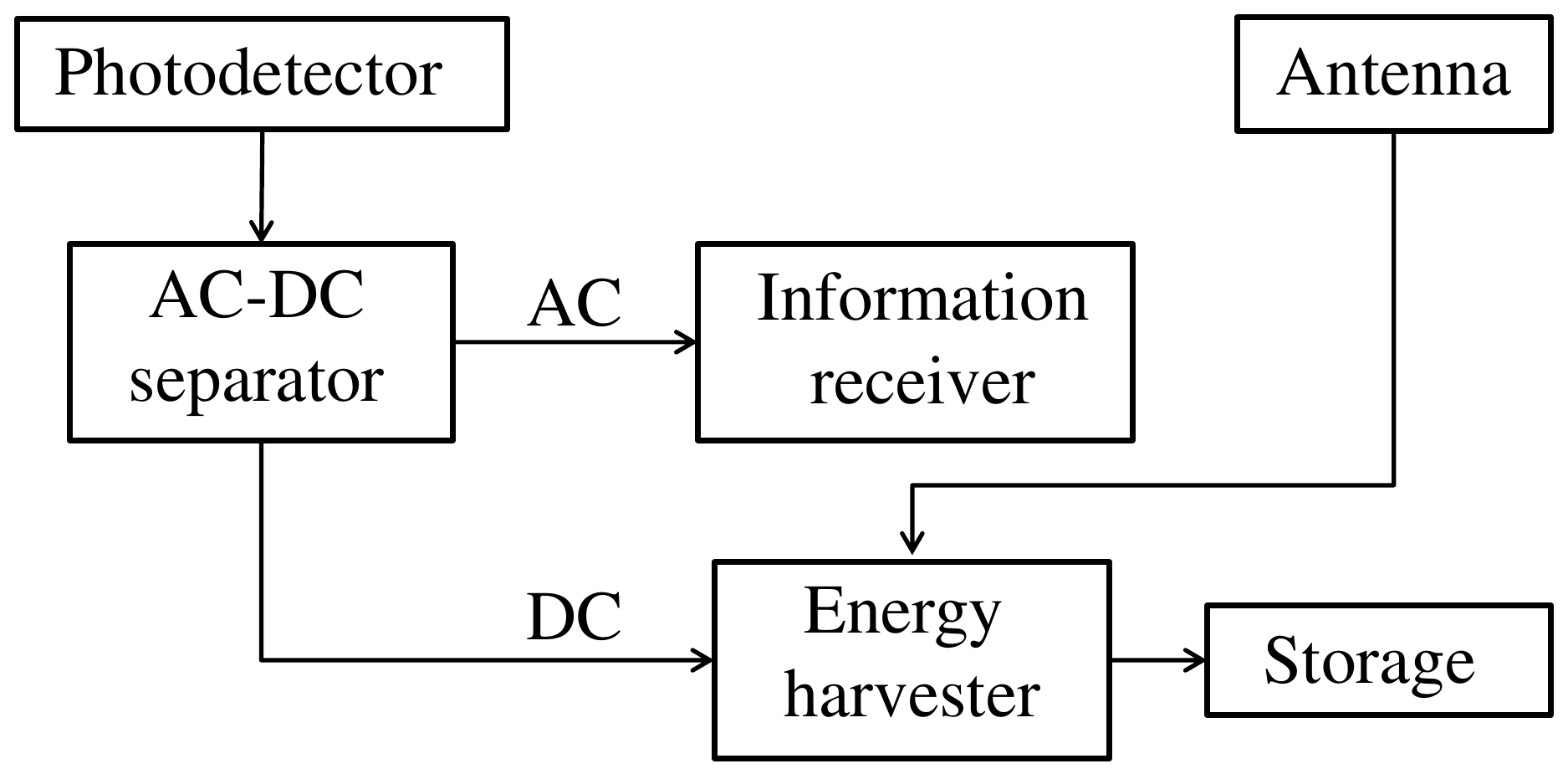}}
   \caption{
      The receiver structure of the device with a multi-homing capability, inspired by\cite{Kashef2016}.
    }
    \label{fig:SystemModel3}
\end{figure}

We consider a receiver structure of  the device with a multi-homing capability, as shown in Fig. \ref{fig:SystemModel3}. 

\subsubsection{Visible Light Communication}
Considering the transmitted optical signals, one of the three colors in ${\mathbb S} = \{ \text{Blue, Green, Red}\}$, denoted by $\mathtt{s} \in {\mathbb S}$, is selected to convey communication data whereas the others, denoted by $\mathtt{c} \in \mathbb {\bar S}$ $({\mathbb {\bar S}} = {\mathbb S} \backslash \{\mathtt{s}\} )$, are used to keep illumination constant.
The optical signal from LED element $i$ of optical transmitter $o$ to device $j$ can be expressed as
\begin{align}\label{eq:VLCsig}
x_{oi,j} &= N_{\mathtt{s}}V_{\mathtt{s}} [A^{\mathtt{s}}_{oi,j} s_{oi,j} + B^{\mathtt{s}}_{oi,j} ] h_{oi,j} \nonumber \\
&+ \sum\limits_{\mathtt{c} \in {\mathbb {\bar S}}} N_{\mathtt{c}}V_{\mathtt{c}} B^{\mathtt{c}}_{oi,j}h_{oi,j}, 
\end{align}
where $N_{\mathtt{s}}$ and $N_{\mathtt{c}}$ are the numbers of LEDs at each element dedicated for the selected and other colors, respectively. $V_{\mathtt{s}}$ and $V_{\mathtt{c}} $ are the LED voltages of the selected and the other colors, respectively. Additionally, $s_{oi,j}$ denotes the modulated electrical signal with zero mean and a unity variance \cite{ChengChen2016}, $A^{\mathtt{s}}_{oi,j}$ is the AC component of the selected color associated with $s_{oi,j}$, $B^{\mathtt{s}}_{oi,j}$ and $B^{\mathtt{c}}_{oi,j}$ represent the DC bias components used for the selected and the other colors respectively. In this concern, $B^{\mathtt{s}}_{oi,j} \in [I_L,I_H]$ and $B^{\mathtt{c}}_{oi,j} \in [I_L,I_H]$, where $[I_L,I_H]$ are the minimum and the maximum input bias currents. Further, to mitigate the effect of clipping distortion, the limitation applied on $A^{\mathtt{s}}_{oi,j}$ can be shown as
\begin{align}\label{eq:restric}
A^{\mathtt{s}}_{oi,j} \le \text{min}\left\{ B^{\mathtt{s}}_{oi,j} - I_L, I_H - B^{\mathtt{s}}_{oi,j} \right\}.
\end{align}

To ensure white illumination, the average light intensity of the three colors should be the same
\begin{align}\label{eq:condi0}
N_{\mathtt{s}}V_{\mathtt{s}} (A^{\mathtt{s}}_{ij}  {\mathbb E} \left[s_{oi,j}\right] + B^{\mathtt{s}}_{oi,j}) =  N_{\mathtt{c}} V_{\mathtt{c}} B^{\mathtt{c}}_{oi,j} \nonumber \\ \quad
{(\forall \mathtt{c} \in {\mathbb {\bar S}} }, \forall o, \forall i, \forall j). 
\end{align}
 Particularly, since ${\mathbb E} \left[ s_{oi,j}\right] = 0$ \cite{ChengChen2016}, the intensity is determined by the DC component, therefore we have the following condition
\begin{align}\label{eq:condi}
N_{\mathtt{s}}V_{\mathtt{s}} B^{\mathtt{s}}_{oi,j} =  N_{\mathtt{c}} V_{\mathtt{c}} B^{\mathtt{c}}_{oi,j} \quad
{(\forall \mathtt{c} \in {\mathbb {\bar S}}}, \forall o, \forall i, \forall j). 
\end{align}
In particular, for convenience, we generalize $N_{\mathtt{s}} = \{N_{\mathtt{c}}\}=N_{LED}$ $(\forall \mathtt{c} \in {\mathbb {\bar S}})$, and $V_{\mathtt{s}} = \{V_{\mathtt{c}}\}=V_{LED}$ $(\forall \mathtt{c} \in {\mathbb {\bar S}})$. Thus, condition \eqref{eq:condi} can be simplified as
\begin{align}\label{eq:condi2}
 B^{\mathtt{s}}_{oi,j} =  B^{\mathtt{c}}_{oi,j} = B, \quad
( \forall \mathtt{c} \in {\mathbb {\bar S}}, \forall o, \forall i, \forall j).
\end{align}

Since the AC component is used for the ID, the SNR expression can be derived as
\begin{align}\label{eq:VLCSNR}
\mathtt {SNR}^{\mathtt{s}}_{oi,j} = \frac{ \left(  \nu N_{LED} V_{LED} h_{oi,j} A^{\mathtt{s}}_{oi,j} \right)^2} {  \sigma^2 },
\end{align}
where $\nu$ is photodetector responsivity and $\sigma^2$ is the noise power.

\subsubsection{VLC Energy Harvesting}
It is obvious that the VLC signals include the DC component, which is separated and then conveyed to the energy harvester, as shown in Fig. \ref{fig:SystemModel3}.
The VLC harvested energy at device $j$ can be computed as \cite{Diamantoulakis2018}
\begin{align}\label{eq:VLCEH}
 \mathtt {EH}_{j}^{VLC} = fI_{j,G} V_{j,c},
\end{align}
where $f$ is the fill factor \cite{CLi2011}, and $I_{j,G}$ is the light generated current. Particularly, since the three colored lights contribute to the EH performance and a device can benefit from multiple light sources, ${I_{j,G}}$ can be calculated as
\begin{align}\label{eq:IG}
I_{j,G} = 3 \nu N_{LED} V_{LED} B \sum\limits_{o}^{\mathcal O} \sum\limits_{i}^{M_I} h_{oi,j},
\end{align}
and $V_{j,c}$ is the open circuit voltage, computed as
\begin{align}\label{eq:Voc}
V_{j,c} = V_0 \text{ln} \left(1 + \dfrac{I_{j,G}}{I_D}\right),
\end{align}
where $V_0$ is the thermal voltage and $I_D$ is the dark saturation current.

\subsubsection{RF Wireless Power Transfer and Energy Harvesting}
On the other hand, RF signals beamed to device $j$ can be given by 
\begin{align}
x^{RF}_{j} = s^{RF}_{j} {\mathbf g}_{j}^H {\mathbf w}_{j},
\end{align}
where $s^{RF}_{j}$ is the unit energy signal and ${\mathbf w}_{j} \in \mathbb C^{M_T \times 1}$ is the beamforming vector.

Thus, the RF harvested energy input at device $j$ can be expressed as
\begin{align}
 \mathtt {\hat EH}_j^{RF} = \sum\limits _{j'=1}^{J} \left| {\mathbf g}_{j}^H {\mathbf w}_{j'} \right|^2.
\end{align}

Considering conventional linear RF EH models, the actual harvested energy, denoted by $\mathtt {EH}_j^{RF}$, can be computed as a linear function of the input energy, i.e. $\mathtt {EH}_j^{RF} = \xi_j^{RF} \mathtt {\hat EH}_j^{RF}$ in which $\xi_j^{RF}$ $(0 \le \xi_j^{RF} \le 1)$ is the energy conversion efficiency. Nevertheless, in practice, the RF energy conversion efficiency is not a constant, its value depends on the strength of the input energy. Hence, a practical non-linear RF EH model is considered in this work. Based on \cite{Clerckx2016,Boshkovska2015a,Boshkovska2017,RuihongJiang2017,KeXiong2017,MengLinKu2018, Tabassum2015}, the non-linear model is
\begin{align}\label{eq:PCEH2}
\mathtt {EH}_j^{RF}= \dfrac{ \dfrac{\mathtt{M}^{EH}}{1 + e^{-\mathtt{a}(\mathtt {\hat EH}_j^{RF} -\mathtt{b})} } - \dfrac{\mathtt{M}^{EH}}{1 + e^\mathtt{ab}} }{1 - \dfrac{1}{1 + e^\mathtt{ab}}},
\end{align}
where $\mathtt{M}^{EH}$ is a positive constant representing the maximum harvested energy at a user when the RF EH circuit meets saturation. In addition, $\mathtt{a}$ and $\mathtt{b}$ are positive constants related to the circuit's specifications, e.g. the resistance, capacitance, and diode turn-on voltage \cite{Clerckx2016,Boshkovska2015a,Boshkovska2017,RuihongJiang2017,KeXiong2017}.

\section{Problem Formulation }
In this section, we explain the aim of the proposed resource allocation scheme and then show the corresponding problem formulation.

In VLC systems, there are some major limitations on the light EH performance. First, indoor light EH might be limited, compared with outdoor solar light EH \cite{Mathews}. Second, the light EH performance might not be stable during non-working times, such as at night. Third, at the transmitter side, the transmit power and the light beam of the LED elements might not be flexibly changed due to maintaining consistent illumination, a very important criteria. These limitations lead to the fact that the VLC network can fail in managing a high information performance whereas ensuring the EH requirement. 

Thus, we propose the combination of optical and RF WPTs to overcome these drawbacks. In this regard, the group of optical transmitters aims to (i) maximize the minimum of the VLC SNRs under a total EH constraint at each user.
Meanwhile, the RF AP intends to (ii) minimize its transmit power while contributing a certain amount of RF EH performance to the overall network to help improve the achievable VLC SNR performance. 
In this work, since our main goal is to improve the VLC performance, we prioritize  objective (i) over objective (ii).
Therefore, we propose a scheme in which, first, the group of optical transmitters actively allocates resources to achieve objective (i) without considering objective (ii) and then decides the amount of RF EH contributed by the RF AP. Second, the RF AP should allocate RF energy to each user according to the RF EH requirement.

To achieve this, there should be a central processing unit in the hybrid RF/VLC network to handle the scheme. In the first step, the optical transmitters convey the information of $\{h_{oi,j}\}$ to the central unit, responsible of  tackling the problem for the group of the optical transmitters, namely OP$_1$, given as
\begin{subequations}\label{eq:maxVLCSINR}
\begin{align}
	\text{OP$_1$:}  \nonumber \\ 
	 \underset{ \{B, A^{\mathtt{s}}_{oi,j}, \mathtt {EH}_{j}^{RF} \}}  \max & \min \hspace{2pt} \frac{ \alpha_{oi,j} \left(  \nu N_{LED} V_{LED} {h_{oi,j}} A^{{\mathtt{s}}}_{oi,j} \right)^2} {  \sigma^2 },  \label{eq:maxVLCSINRa} 
\\
\text{s.t.:}  	\quad & \mathtt {EH}_{j}^{VLC} + \mathtt {EH}_{j}^{RF} \ge \theta, \quad (\forall j) \label{eq:maxVLCSINRb}			
\\
			\quad &   B^{\mathtt{s}}_{oi,j} = B, \quad ( \forall o, \forall i, \forall j)  \label{eq:maxVLCSINRc}
\\			
			\quad &   A^{\mathtt{s}}_{oi,j}  \le I_H - B^{\mathtt{s}}_{oi,j}, \quad ( \forall o, \forall i, \forall j)  \label{eq:maxVLCSINRc1}
\\
			\quad &   \frac{I_L + I_H}{2} \le B \le I_H,  \label{eq:maxVLCSINRc2}
\\
			\quad & \mathtt {EH}_{j}^{RF} \le \theta^{RF}, \quad (\forall j) \label{eq:maxVLCSINRd}
\end{align}
\end{subequations}
where $\{\alpha_{oi,j}\} \in \{0;1\}$ with $\alpha_{oi,j} = 1$ indicating that LED element $i$ of optical transmitter $o$ serves user $j$ which are known as a priori, and these values are determined by the operator according to the users' geographical position. Further, $\theta$ is the preset threshold of the total EH.
Constraint \eqref{eq:maxVLCSINRc} ensures that the transmit power of the three colors is the same for the white illumination. Further, the restriction on $A^{\mathtt{s}}_{oi,j}$, given in \eqref{eq:restric}, implies that there are two possible values of $B^{\mathtt{s}}_{oi,j}$ resulting in the same $A^{\mathtt{s}}_{oi,j}$. These two values fall in two ranges, $I_L \le B^{\mathtt{s}}_{oi,j} < \frac{I_L + I_H}{2}$ and $\frac{I_L + I_H}{2} \le B^{\mathtt{s}}_{oi,j} \le I_H$, respectively. As mentioned earlier, the indoor light EH in its nature is limited. Since we aim to exploit the light energy and maintain a reasonable illumination in the covered area, we only consider the values of $B$ in the higher range (i.e. $\frac{I_L + I_H}{2} \le B^{\mathtt{s}}_{oi,j} \le I_H$) as feasible solutions. Thus, constraints \eqref{eq:maxVLCSINRc1} and \eqref{eq:maxVLCSINRc2} are formulated. Moreover, $\theta^{RF}$ is the threshold of the RF EH that the RF AP contributes to each user. In this context, it is well-known that the efforts to convey energy wirelessly might lead to RF pollution, resulting in human health issues. There is a restriction on the RF absorption rate applied to the human body \cite{Vuwcnc,IEEEstandard}. Accordingly, this issue is also imposed through \eqref{eq:maxVLCSINRd}. 

\begin{figure}[!]
\centering
{\includegraphics[width=0.4\textwidth]{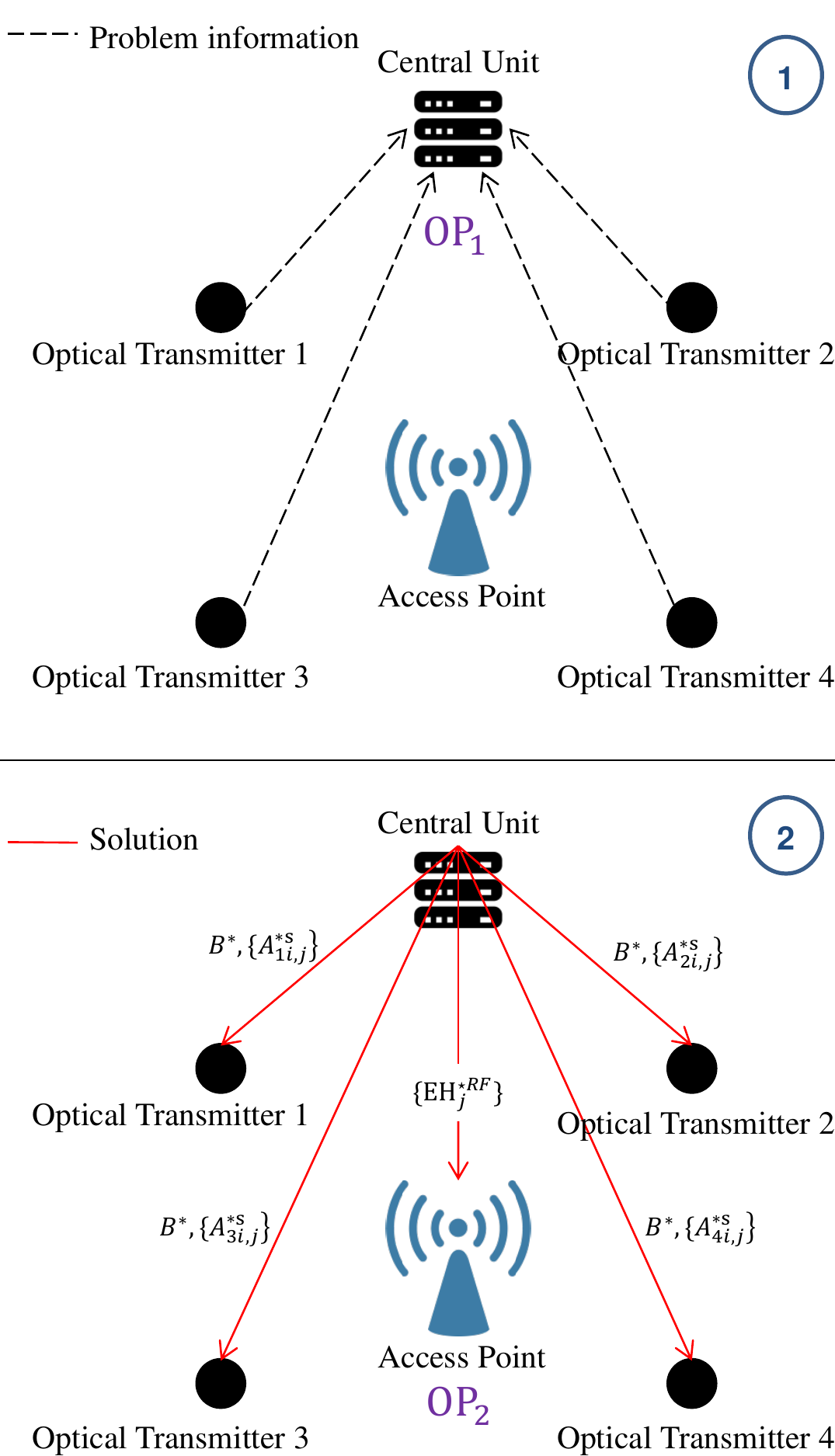}}
   \caption{
    The scenario of solving problems OP$_1$ and OP$_2$.
    }
    \label{fig:centralize}
\end{figure}

\begin{remark}
The solution of OP$_1$, which is optimal in terms of the initial objective (i.e., max-min SNR), also maximizes the VLC SNR for any user in the network. This is because an increase of the SNR for any user directly leads to higher SNRs for all users (due to equal DC component for each LED element). On this basis, choosing the max-min SNR objective for managing QoS implies maximizing the sum and the minimum achievable throughput of the network, with the latter being an increasing function of the SNR for each user \cite{Lapidoth}. This is an important observation, since these objectives are the most commonly used, when the channel knowledge is available at the transmitter.
\end{remark}

Next, in our scheme, the central unit tackles problem OP$_1$ and then distributes the optimal values, $B^{\star}$ and $\{A^{\star{\mathtt{s}}}_{oi,j}\}$, to each optical transmitter while conveying the optimal values $\{ \mathtt {EH}_{j}^{\star RF} \}$ to the RF AP. Therefore, the RF AP is responsible of allocating the energy resources such that the RF EH at each user achieves $\{ \mathtt {EH}_{j}^{\star RF} \}$ obtained from OP$_1$. The optimization problem of the RF AP, namely OP$_2$, is 
\begin{subequations}\label{eq:minpower}
\begin{align}
	\text{OP$_2$:} \quad \underset{ {{\mathbf w}_j}  }  \min  \quad & \sum_{j=1}^{J} \left\| {{\mathbf w}_j} \right\|^2 \quad   \label{eq:minpower} 
\\
\text{s.t.:} \quad & \mathtt {EH}_{j}^{RF} \ge \mathtt {EH}_{j}^{\star RF}, \quad (\forall j) \label{eq:minpowera}
\end{align}
\end{subequations}
where $\mathtt {EH}_{j}^{\star RF}$ is the optimal value of $\mathtt {EH}_{j}^{ RF}$, obtained from solving problem OP$_1$. For convenience, Fig. \ref{fig:centralize} shows that how problems OP$_1$ and OP$_2$ are handled.

\begin{remark}
In realistic scenarios, when the network needs to update solutions, the network operator can exclude $B$ from re-solving OP$_1$ and include it again whenever necessary (i.e., when it is suitable to change the illumination level to increase/decrease system performance). This helps to actively maintain the constant lighting level in covered areas. This further highlights the crucial role of RF AP in fulfilling performance requirements under the changes of the network.
\end{remark}

In general, it is difficult to solve problem OP$_1$. In this regard, considering the objective function, the maximization of a convex function is not a convex problem. Further, constraint \eqref{eq:maxVLCSINRb} is nonconvex since it can be seen that $\mathtt {EH}_{j}^{VLC}$ is concave while $\mathtt {EH}_{j}^{RF}$ is convex (i.e. the second-order condition \cite{StephenBoyd2004}). Therefore, problem OP$_1$ is nonconvex. Furthermore, problem OP$_2$ is nonconvex because constraint \eqref{eq:minpowera} has the form of a convex function larger than a constant. To overcome these issues, we propose novel methods to solve these problems optimally and suboptimally, presented in the next section.

\section{Proposed Solutions}
This section presents our methods to tackle problems OP$_1$ and OP$_2$. First, we propose a method to decompose problem OP$_1$ without the loss of optimality and then solve the corresponding subproblems. Second, we tackle problem OP$_2$. Lastly, we derive a suboptimal solution with low complexity and then introduce a semi-decentralized approach to handle OP$_1$ and OP$_2$.

\subsection{Decomposing Problem OP$_1$ without Loss of Optimality}
In light of {\it Remark} 1, it is worth emphasizing that since all the DC components of the LED elements are the same, increasing the VLC SNR for any user leads to increasing the VLC SNR for all users. On this basis, considering problem OP$_1$, there is an interesting characteristic, given in {\it Proposition} 1 below

\begin{proposition}
Constraint \eqref{eq:maxVLCSINRb} implies that, to improve the VLC SNR, the contribution of VLC EH to the overall EH performance should be lowered through increasing RF EH. 
Since maximizing the minimum of VLC SNRs is equivalent to maximizing the VLC SNR at the user with the worst VLC SNR, the  RF EH allocated to the user with the worst VLC SNR determines the value of $B$ for all the users. 
\end{proposition}

\begin{IEEEproof}
See Appendix A.
\end{IEEEproof}

Based on {\it Proposition} 1, we propose an approach to decompose problem OP$_1$ into two sub-problems without loss of optimality. 
The resource allocation can be divided into two stages: 
First, considering user $\bar j$ (where ${\bar j}$ is the user with the worst VLC SNR), the optimal values of $\{\mathtt {EH}_{\bar j}^{RF}\}$ denoted by $\{\mathtt {EH}_{\bar j}^{\star RF}\}$, need to be determined. Second, by replacing $\{\mathtt {EH}_{\bar j}^{RF}\}$ by $\{\mathtt {EH}_{\bar j}^{\star RF}\}$ in OP$_1$, the optimal values of $\{A^{\textcolor{red}{\mathtt{s}}}_{oi,j}\}$ and $B$ are then found. It is clear that the optimality of this decomposition is preserved.

To excute the first stage, we omit the parts unrelated with $\{\mathtt {EH}_{\bar j}^{RF}\}$ in OP$_1$. Accordingly, the corresponding problem, namely SubRF, can be formulated as
\begin{subequations}\label{eq:RFsubProblem0}
\begin{align}
	\text{SubRF:} 
\underset{ \{B, A^{{\mathtt{s}}}_{oi,j},\mathtt {EH}_{\bar j}^{RF}\}}  \max &  \quad \hspace{2pt} \frac{ \left(  \nu N_{LED} V_{LED} {h_{oi,\bar j}} A^{{\mathtt{s}}}_{oi,\bar j} \right)^2} {  \sigma^2 },  \quad  \label{eq:RFsubProblema} 
\\
\text{s.t.:}  	\quad &  \mathtt {EH}_{\bar j}^{VLC} + \mathtt {EH}_{\bar j}^{RF} \ge \theta \label{eq:RFsubProblemb}
\\
			\quad &\mathtt {EH}_{\bar j}^{RF} \le \theta^{RF},   \label{eq:RFsubProblemb}
\end{align}
\end{subequations}

In the second stage, we substitute $\{\mathtt {EH}_{\bar j}^{\star RF}\}$ into OP$_1$. The resulting problem, namely SubVLC, can be written as
\begin{subequations}\label{eq:VLCsubProblem0}
\begin{align}
	\text{SubVLC:} 
\underset{ B, A^{\mathtt{s}}_{oi,j},  \mathtt {EH}_{j}^{RF}}  \max &  \min \hspace{2pt} \frac{ \alpha_{oi,j} \left(  \nu N_{LED} V_{LED} {h_{oi,j}} A^{{\mathtt{s}}}_{oi,j} \right)^2} {  \sigma^2 },  \quad  \label{eq:VLCsubProblema} 
\\
\text{s.t.:}  	\quad & \mathtt {EH}_{{\bar j}}^{VLC} \ge \theta -  \mathtt {EH}_{{\bar j}}^{\star RF},  \label{eq:VLCsubProblemb}			
\\			
			\quad & \mathtt {EH}_{{ j}}^{VLC} \ge \theta -  \mathtt {EH}_{j}^{ RF},  (\forall j, j \neq \bar j ) \label{eq:VLCsubProblemc}			
\\			
			\quad &   A^{\mathtt{s}}_{oi,{ j}}    \le I_H - B,   ( \forall o, \forall i, \forall j) \label{eq:VLCsubProblemd}
\\
			\quad &   \frac{I_L + I_H}{2} \le B \le I_H.  \label{eq:VLCsubProbleme}
\end{align}
\end{subequations}

Generally, since the principle of optimally allocating the resources is unveiled through {\it Remark} $1$ and {\it Proposition} $1$, it is obvious that $\mathtt {EH}_{\bar j}^{\star RF}$, the optimal solution of problem SubRF, is also the optimal one of problem OP$_1$. Thus, the optimality of our method is guaranteed. 

\subsection{Solution to Sub-problem SubRF}
To solve problem SubRF, we start with finding the feasible range of $\mathtt {EH}_{\bar j}^{ VLC}$ through its maximum and minimum, denoted as $\{\max \mathtt {EH}_{\bar j}^{ VLC}\}$ and $\{\min \mathtt {EH}_{\bar j}^{ VLC}\}$, respectively.
It is clear that, according to \eqref{eq:condi2}, $\mathtt {EH}_{\bar j}^{ VLC}$ in problem SubRF achieves its maximum when $\{A^{\mathtt{s}}_{oi,{\bar j}}\}=0$ and $ B = I_H$. Thus, we have
\begin{align}
\max \mathtt {EH}_{\bar j}^{VLC} = f\hat I_{{\bar j}G} V_0 \text{ln} (1 + \dfrac{\hat I_{{\bar j},G}}{I_D}),
\end{align}
where
\begin{align}
\hat I_{{\bar j},G} &= 3\nu N_{LED}V_{LED} I_H \sum\limits_{o}^{\mathcal O} \sum\limits_{i}^{M_I} h_{oi,{\bar j}}.
\end{align}

Next, $\mathtt {EH}_{\bar j}^{ VLC}$ in problem SubRF achieves the minimum when $\{A^{\mathtt{s}}_{oi,{\bar j}}\}$ reach the maximum, i.e., \eqref{eq:restric}.
In more details, the maximum of $\{A^{\mathtt{s}}_{oi,{\bar j}}\}$ and the corresponding values of $B$ in problem SubRF can be computed, respectively, as follows
\begin{align}\label{eq:A}
\hat A^{\mathtt{s}}_{oi,{\bar j}} = \dfrac{I_H + I_L}{2}, 
\end{align}
\begin{align}\label{eq:B}
\bar B=   \dfrac{I_H - I_L}{2}. 
\end{align}
Based on \eqref{eq:A} and \eqref{eq:B}, $\min \mathtt {EH}_{\bar j}^{ VLC}$ can be given by 
\begin{align}
\min \mathtt {EH}_{\bar j}^{VLC} = f\bar I_{{\bar j},G} V_0 \text{ln} (1 + \dfrac{\bar I_{{\bar j},G}}{I_D}),
\end{align}
where
\begin{align}
\bar I_{{\bar j},G} &= 3 \nu   N_{LED}V_{LED}\bar B \sum\limits_{o}^{\mathcal O} \sum\limits_{i}^{M_I} h_{oi,{\bar j}}.
\end{align}

In general, the feasibility of problem SubRF can be confirmed using the condition below. Problem SubRF is feasible if and only if
\begin{align}
\theta - \max \mathtt {EH}_{\bar j}^{VLC} \le \theta^{RF}.
\end{align}

After some mathematical calculations, the optimal solution can be computed as follow
\begin{align}
\mathtt {EH}_{\bar j}^{\star RF} = \min \{ \theta - \min \mathtt {EH}_{\bar j}^{VLC}, \theta^{RF} \}.
\end{align}

\subsection{Optimal Solution to Sub-problems SubVLC}
One can observe that sub-problem SubVLC is subject to the class of maximizing a convex function. Conventionally, the sub-problem can be solved by applying successive convex approximation (SCA) methods \cite{Beck2010}. However, this approach can bring a significant computational burden. In this work, we show a method to transform the sub-problem into a convex formulation by exploiting the sub-problem's characteristics.

According to {\it Proposition} $1$, solving SubVLC can be achieved by tackling two alternative sub-problems, namely SubVLC$_{\bar j}$ and SubVLC$_{j}$, with respect to user $\bar j$ and the other users, respectively. Thus, the optimization problem associated with user $\bar j$ can be formulated as
\begin{subequations}\label{eq:VLCsubProblem1}
\begin{align}
	\text{SubVLC$_{\bar j}$:}  \underset{ B, A^{\mathtt{s}}_{oi,{\bar j}}}  \max &  \quad  \frac{ \left(  \nu N_{LED} V_{LED} {h_{oi,\bar j}} A^{{\mathtt{s}}}_{oi,\bar j} \right)^2} {  \sigma^2 }  \quad  \label{eq:VLCsubProblem1a} 
\\
\text{s.t.:}  	\quad & \mathtt {EH}_{\bar j}^{VLC} \ge \theta-  \mathtt {EH}_{\bar j}^{\star RF},  \label{eq:VLCsubProblem1b}			
\\			
			\quad &   A^{\mathtt{s}}_{oi,{\bar j}}    \le I_H - B,  \label{eq:VLCsubProblem1d}
\\
			\quad &   \frac{I_L + I_H}{2} \le B \le I_H.  \label{eq:VLCsubProbleme}
\end{align}
\end{subequations}
Further, the optimization problem with respect to the other users $(\forall j, j \neq \bar j )$ is
\begin{subequations}\label{eq:VLCsubProblem11}
\begin{align}
	\text{SubVLC$_j$:} \quad \underset{ A^{\mathtt{s}}_{oi,j}}  \max  \quad  & \frac{ \alpha_{oi,j} \left(  \nu N_{LED} V_{LED} {h_{oi,j}} A^{{\mathtt{s}}}_{oi,j} \right)^2} {  \sigma^2 }  \quad  \label{eq:VLCsubProblem11a} 
\\
\text{s.t.:} 
			\quad &   A^{\mathtt{s}}_{oi,j}    \le I_H - B^{\star}. \label{eq:VLCsubProblem11b}
\end{align}
\end{subequations}
It is worth noting that $B^{\star}$ is the optimal value of $B$, obtained from solving sub-problem SubVLC${\bar j}$. Also, we obtain
\begin{align}\label{eq:EHopt}
\mathtt {EH}_{j}^{\star RF} = \theta - \mathtt {EH}_{j}^{VLC } (B^{\star}).
\end{align}

We start with solving SubVLC$_{\bar j}$. According to \eqref{eq:VLCSNR}, maximizing $\mathtt {SNR}^{\mathtt{s}}_{oi,{\bar j}}$ can be achieved by maximizing $A^{\mathtt{s}}_{oi,{\bar j}}$. Moreover, considering constraints \eqref{eq:VLCsubProblemc} and \eqref{eq:VLCsubProblemd}, one can observe the relationship between $A^{\mathtt{s}}_{oi,{\bar j}}$, and $B$. Constraint \eqref{eq:VLCsubProblemd} always holds with equality at the optimal values of $\{A^{\mathtt{s}}_{oi,{\bar j}}\}$. Hence, maximizing $A^{\mathtt{s}}_{oi,{\bar j}}$ is equivalent to minimizing $B$. In light of this discussion, solving problem SubVLC$_{\bar j}$ can be equivalently reformulated into two steps. First, the optimal solution of $B$, denoted by $B^{\star}$, in problem SubVLC$_{\bar j}$ can be found by solving the problem below
\begin{subequations}\label{eq:VLCsubProblem2}
\begin{align}
	  \underset{  B}  \min \quad &  B  \quad  \label{eq:VLCsubProblem2a} 
\\
\text{s.t.:}  	
			\quad & \mathtt {EH}_{\bar j}^{VLC} \ge \theta_j -  \mathtt {EH}_{\bar j}^{\star RF},  \label{eq:VLCsubProblem2c}	\\			
			\quad &  \dfrac{I_H - I_L}{2} \le B \le I_H.
\end{align}
\end{subequations}

Second, the optimal value of $A^{\mathtt{s}}_{oi,{\bar j}}$ in sub-problem SubVLC$_{\bar j}$ can be given by
\begin{align}\label{eq:Aop1}
A^{\star \mathtt{s}}_{oi,{\bar j}} = I_H - B^{\star}.
\end{align}

Following \eqref{eq:VLCEH} and \eqref{eq:IG}, constraint \eqref{eq:VLCsubProblem2c} needs to be represented into a more tractable form, given by
\begin{align}\label{eq:VLCsubProblemb1}
 \text{ln} \left(1 + \dfrac{I_{{\bar j},G} (B)}{I_D}\right) \ge \frac{\theta_{\bar j} -  \mathtt {EH}_{\bar j}^{\star RF}}{f V_0 I_{{\bar j},G} (B) },
\end{align}
where, for convenience, $I_{{\bar j},G} (B)$ denotes that $I_{{\bar j},G}$ is a function of $B$. Accordingly, replacing \eqref{eq:VLCsubProblem2c} by \eqref{eq:VLCsubProblemb1}, problem \eqref{eq:VLCsubProblem2} becomes convex. 
In particular, solving problem \eqref{eq:VLCsubProblem2} can be achieved by a bisection-based algorithm as follows
\begin{center}
\begin{tabular}{|l|}
\hline
\textbf{Algorithm 1}: Algorithm to find $B^{\star}$ \tabularnewline
\hline
\hline
i. Set $B_{min} =  \dfrac{I_H - I_L}{2}$, $B_{max} = I_H$, and a solution \tabularnewline
 accuracy $\varsigma \ge 0$.\tabularnewline
{\bf Repeat} \tabularnewline
ii. {\it If} constraint \eqref{eq:VLCsubProblemb1} is satisfied with $B = \frac{B_{min} + B_{max}}{2} $ \tabularnewline
\hspace{25pt}{\it then} $B_{min} =B$ \tabularnewline
iii. {\it Else} $B_{max} =B$. \tabularnewline
{\bf Until} $B_{max} - B_{min} \le \varsigma$. \tabularnewline
iv. The output $B$ is the desired solution. \tabularnewline
\hline
\end{tabular}
\end{center}

\begin{figure}[t]
\centering
{\includegraphics[width=0.3\textwidth]{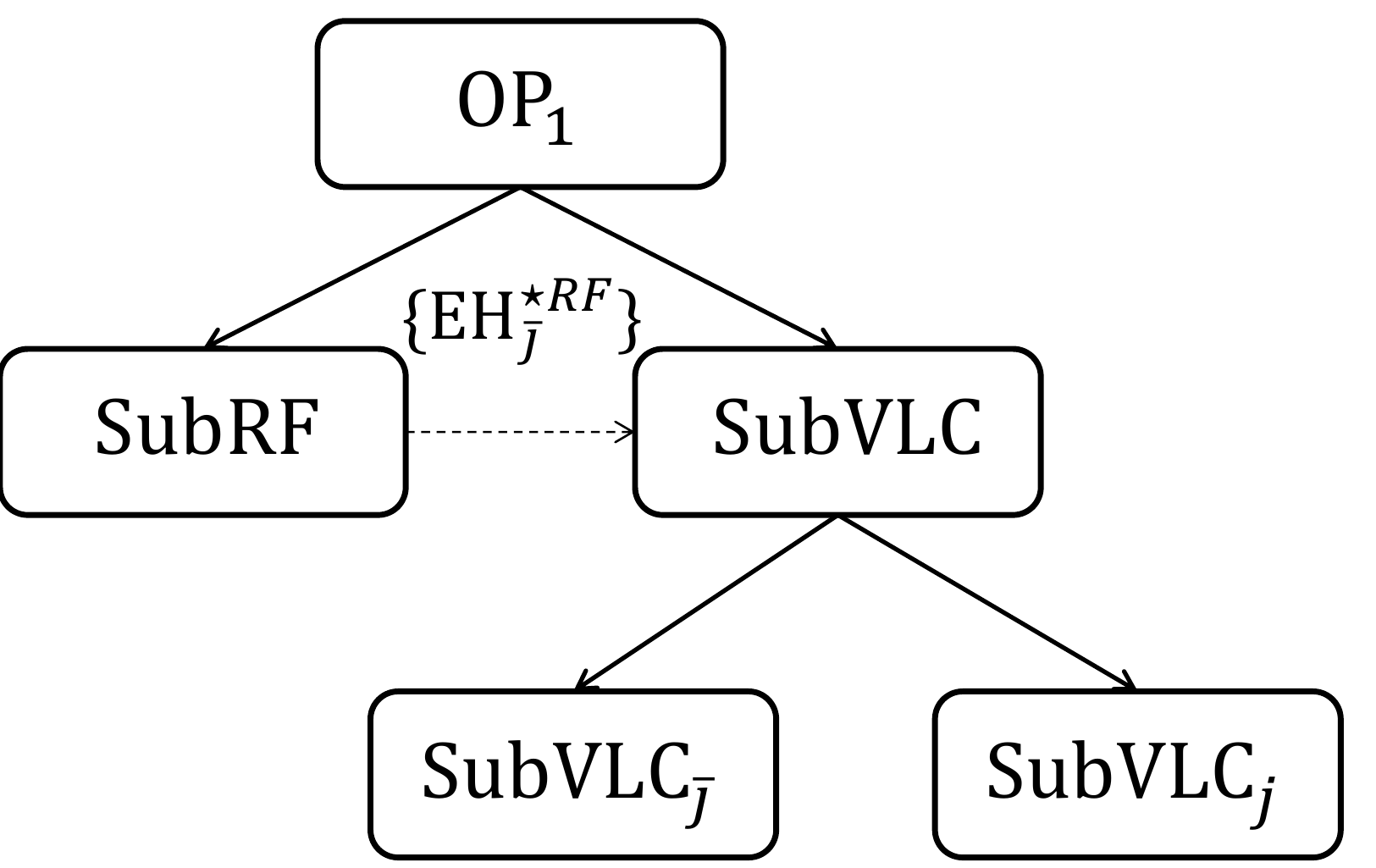}}
   \caption{
    Flowchart of handling OP$_1$.
    }
    \label{fig:OP1}
\end{figure}

After obtaining $B^{\star}$, we consider sub-problem SubVLC$_j$. It can be seen that constraint \eqref{eq:VLCsubProblem1b} holds optimality with equality. Therefore, the optimal solution SubVLC$_j$ can be calculated as
\begin{align}\label{eq:Aop2}
A^{\star \mathtt{s}}_{oi,{j}} = I_H - B^{\star}, \quad (\forall j, j \neq \bar j ).
\end{align}
Combining \eqref{eq:Aop1} and \eqref{eq:Aop2}, for all users, we have
\begin{align}\label{eq:Aop3}
A^{\star \mathtt{s}}_{oi,{j}} = I_H - B^{\star}, \quad (\forall j).
\end{align}

For convenience, a flowchart of handling problem OP$_1$ is shown in Fig. \ref{fig:OP1}.

\subsection{Solution to Problem OP$_2$}
In this subsection, we propose an efficient method to solve OP$_2$. 

To deal with problem OP$_2$, first, we can transform the constraint into a more tractable form based on \cite{RuihongJiang2017}. Next, by defining ${\mathbf W}_j = {\mathbf w}_j {\mathbf w}_j^H$ and ${\mathbf G}_{j} = {\mathbf g}_{j} ({\mathbf g}_{j})^H$, problem OP$_2$ can be equivalently reformulated as 
\begin{subequations}\label{eq:SDRprob}
\begin{align}
	\underset{ \{{\mathbf W}_j\}  }  \min  \quad & \sum_{j=1}^{J}  \text{tr}({{\mathbf W}_j})  \quad  
\\
\text{s.t.:} \quad & \sum\limits _{j'=1}^{J}  \text{tr} ({\mathbf W}_{j'} {\mathbf G}_{j}) \ge \bar \theta_j , \quad (\forall j), \label{eq:SDRproba}
\\
			\quad & {\mathbf W}_j \succeq 0, \quad (\forall j), \label{eq:SDRprobb}
\\
			\quad & \text{Rank} ({\mathbf W}_j) = 1, \quad (\forall j). \label{eq:SDRprobc}
\end{align}
\end{subequations}
in which, from \eqref{eq:PCEH2},
\begin{align}
\bar \theta_j = \mathtt{b} - \frac{1}{\mathtt{ a}}\text{ln}\left(  \frac{e^{\mathtt {ab}}({\mathtt M}^{EH} - \mathtt {EH}_{j}^{\star RF})}{e^\mathtt{ ab}\mathtt {EH}_{j}^{\star RF} + \mathtt M^{EH} } \right).
\end{align}
The constraint $\text{Rank} ({\mathbf W}_j) = 1$ is set to guarantee that ${\mathbf W}_j = {\mathbf w}_j {\mathbf w}_j^H$ holds after optimizing ${\mathbf W}_j$.

Problem OP$_2$ has been transformed into a form of semi-definite programming (SDP). Also, it is still non-convex due to the rank-one constraint. However, we can demonstrate that the rank-one condition always holds in problem OP$_2$ through {\it Lemma} 1 as follows
\begin{lem}
Considering problem \eqref{eq:SDRprob}, the constraint of rank one condition, i.e. $\text{Rank} ({\mathbf W}_j) = 1$  $(\forall j)$, always does hold.
\end{lem}
\begin{IEEEproof}
See Appendix B.
\end{IEEEproof}

Accordingly, we can omit constraint \eqref{eq:SDRprobc} and then tackle \eqref{eq:SDRprob} using SDP solvers, such as CVX \cite{Gra2009}.

\subsection{Suboptimal Solution to SubVLC and Proposed Semi-Decentralized Approach}
In the continuity, to facilitate the solving process and to reduce the computational burden at the central control unit, we present a suboptimal solution to sub-problem SubVLC$_j$ through {\it Lemma} 2 below
\begin{lem}
The closed-form suboptimal solution to problem SubVLC, denoted by $A^{\times \mathtt{s}}_{oi,j}$ and $B^{\times}$, can be calculated by

\begin{align}
B^{\times} &= \min \left\{ \max \{ \Upsilon, I_L\},  I_H \right\}, \label{eq:Btimes} \\
A^{\times \mathtt{s}}_{oi,j} &= I_H - B^{\times}, \quad ( \forall j)
\end{align}
where
\begin{align}
\Upsilon =  \frac{\theta -  \mathtt {EH}_{\bar j}^{\star RF}}{3\nu N_{LED}V_{LED}f V_0 \mathcal W \left( \frac{\theta -  \mathtt {EH}_{\bar j}^{\star RF}}{f V_0 I_D}\right) \sum\limits_{o}^{\mathcal O} \sum\limits_{i}^{M_I}h_{oi,{\bar j}}},
\end{align}
where $\mathcal W (.)$ is the Lambert function.
\end{lem}

\begin{IEEEproof}
See Appendix C.
\end{IEEEproof}

Indeed, the suboptimal solution can be calculated with low complexity. Assuming that each optical transmitter is capable of computing the suboptimal solution of SubVLC$_j$ (i.e., it is equipped with a micro controller), we propose a semi-decentralized approach for the network as follows. 

Instead of solving OP$_1$, the control unit solves only problem SubRF and then distributes the information of $\mathtt {EH}_{\bar j}^{\star RF}$ and $h_{\bar j}$ $\left( h_{\bar j} = \sum\limits_{o}^{\mathcal O} \sum\limits_{i}^{{M_I}}h_{oi,\bar j} \right)$ to the optical transmitters and the RF AP. Afterwards, each optical transmitter tackles its own problem SubVLC$_j$ by using the closed-form expression provided in {\it Lemma} 2. Also, the RF AP solves problem OP$_2$ to get the optimal beamformers.

\begin{figure}[!]
\centering
{\includegraphics[width=0.45\textwidth]{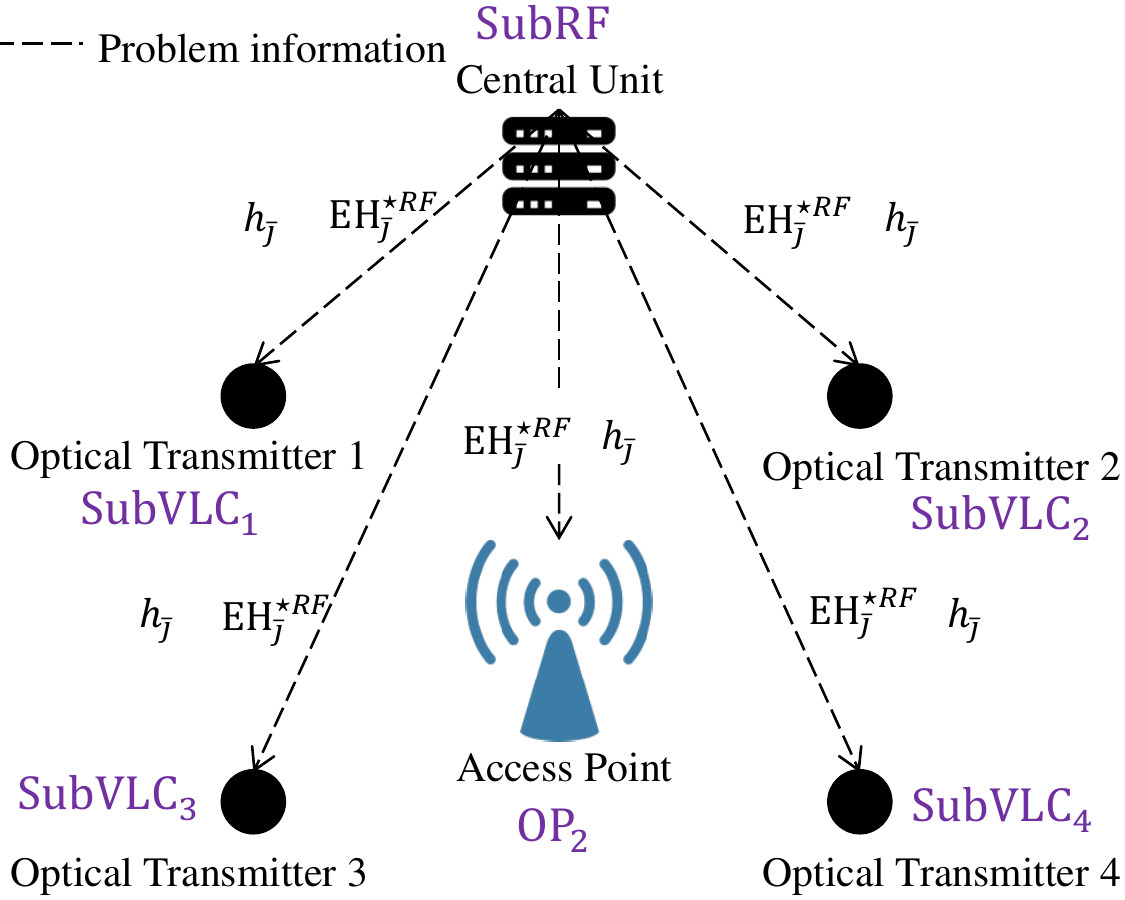}}
   \caption{
      An example of the proposed decentralized scenario.
    }
    \label{fig:decentralize}
\end{figure}

For convenience, our decentralizing method is illustrated in Fig. \ref{fig:decentralize}. The advantages of this approach can be listed as: (i) all the computation burden is no longer put on the control unit but shared between the optical transmitters, (ii) the optical transmitters can adapt to the changes of the networks (i.e. users' position) instantly, and (iii) since the Lambert function $\mathcal W (.)$ is a simple well-known function and is easy to compute within nano seconds using average CPUs \cite{Fukushima2013}, it can be seen that the closed-form suboptimal solution has low complexity. The performance of this approach relies on the sub-optimality of the solution which is verified in the numerical section.

\section{Numerical Results}

Regarding the simulation, we consider a hybrid RF/VLC network consisting of one 6-antenna RF AP, four 7-angle-diversity optical transmitters and five user devices. 
We consider an office environment, as shown in Fig. \ref{fig:SystemModel1}, with a size of $5 \times 5 \times 3$ m$^3$. The locations of the four optical transmitters are $(1.5, 1.5, 3)$, $(1.5, 3.5, 3)$,
$(3.5, 3.5, 3)$, and $(3.5, 1.5, 3)$, respectively. The distance between the transmitters and the receiver plane is $2$ meters. The RF AP is located at $(2.5, 2.5, 3)$.

Regarding VLC channels, based on pioneering works \cite{ZheChen2018,Diamantoulakis2018,TamerRakia,GaofengPan}, we set $T_s(\psi_{oi,j} ) = 1$ , $\phi_{oi,1/2} = 17^\circ$, $n = 1.5$, $\psi_{oi,j,c} = 60^\circ$, $I_L = 2$ mA, $I_H = 12$ mA, $B = I_H$, $I_D = 10^{-9}$ A, $\sigma^2 = 10^{-15}$ . The distances between the five users and their corresponding optical transmitters are 2.05 m, 2.10 m, 2 m, 2.11 m, and 2.08 m, respectively. Particularly, similar to \cite{ZheChen2018}, with $\phi_{oi,1/2} = 17^\circ$ for each LED element, the 7-element angle-diversity transmitters are designed to have a LED semiangle at half-power equal to $60^\circ$ which is equivalent to the one in conventional single-element transmitters. Further, according to Wysips Reflect \cite{wysipscrystal}, a solar pannel can be integrated into the phone screen, thus we can set $Ar_j = 85$ cm$^2$ (i.e. the screen size of a Samsung Galaxy S8). Also, $\nu$ = 0.4, and $f=0.75$. In addtion, we also set the number of LEDs as $N_{LED} = 40$ and the LED voltage as $V_{LED}= 2.25$ V \cite{Larimore}. All the optical transmitters have the same settings.

Additionally, considering the RF channels, we set the Rician factor $R = 6$ dB and the exponent path loss factor to 2.6, suitable for office environments \cite{Rappaport94}. Also, we investigate the system performance for $\theta^{RF} \le 6$ mW which is the limit for human health safety \cite{Vuwcnc}.
Regarding the nonlinear RF EH model, we set $\mathtt{M}^{EH} = 24$ mW, $\mathtt{a} = 150$ and $\mathtt{b} = 0.014$ based on the mathematical analysis of the practical nonlinear RF EH model provided in \cite{Boshkovska2015a, Boshkovska2017,RuihongJiang2017,KeXiong2017}. On this basis, the numerical results are shown and discussed as follows.

\begin{figure}[!]
\centering
{\includegraphics[width=0.45\textwidth]{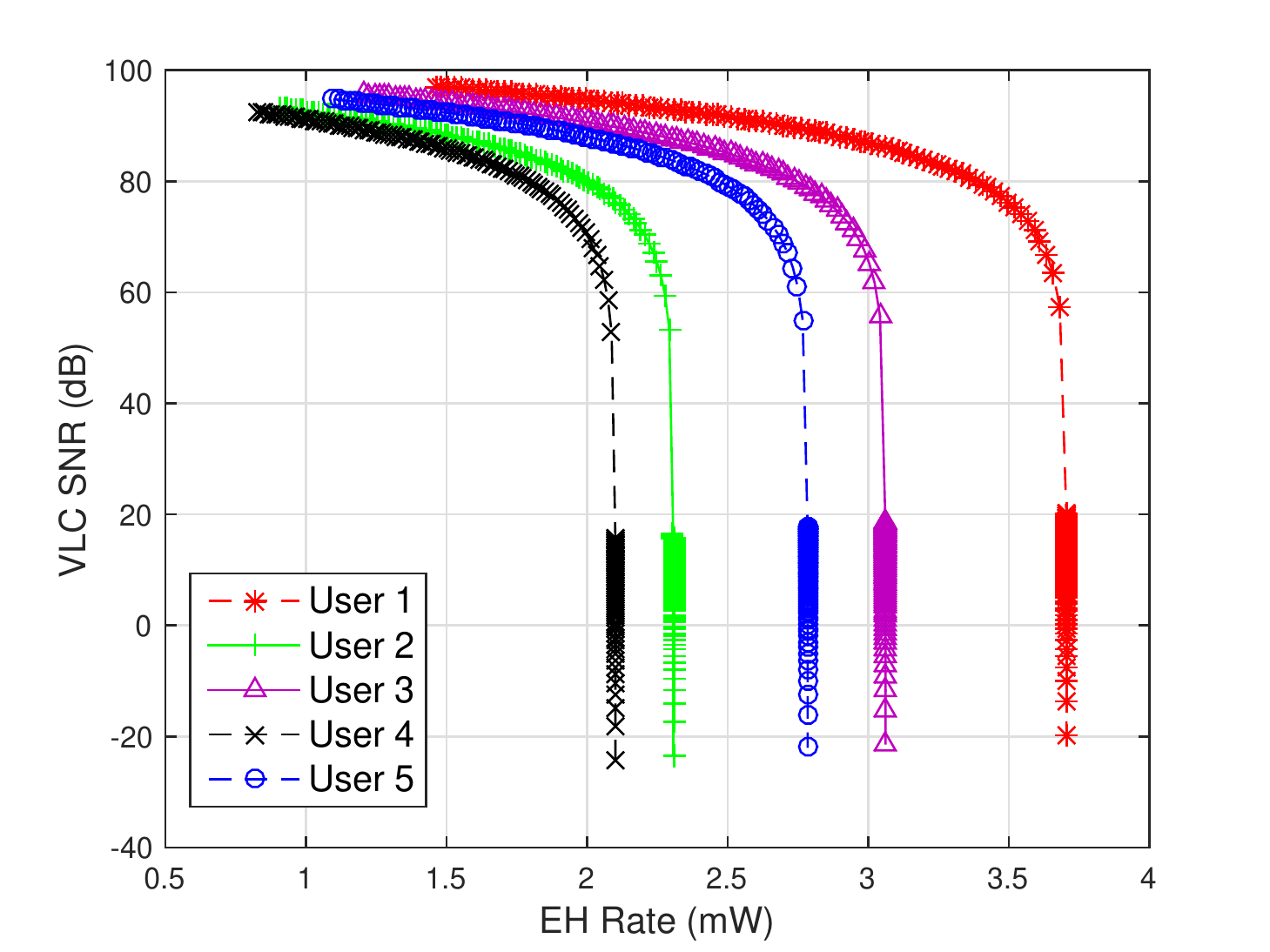}}
   \caption{
      SNR-EH region at each user.
    }
    \label{fig:SNREH}
\end{figure}

Fig. \ref{fig:SNREH} exhibits the VLC SNR-EH regions for all the users without the help of the RF AP. It is obvious that a user with a shorter VLC transmission distance has a larger VLC SNR-EH region. Particularly, as observed, the minimum VLC EH is always larger than zero for all users. This can be explained by the restriction applied on the values of the AC components $\{A^{\mathtt{s}}_{oi,j}\}$, mathematically represented by \eqref{eq:restric}, which implies that the maximum $\{A^{\mathtt{s}}_{oi,j}\}$ is less than $I_L$, and therefore $\{B^{\mathtt{s}}_{oi,j}\}$ is always positive. On this basis, the VLC EH is always greater than zero for all users.

\begin{figure}[t]
\centering
{\includegraphics[width=0.45\textwidth]{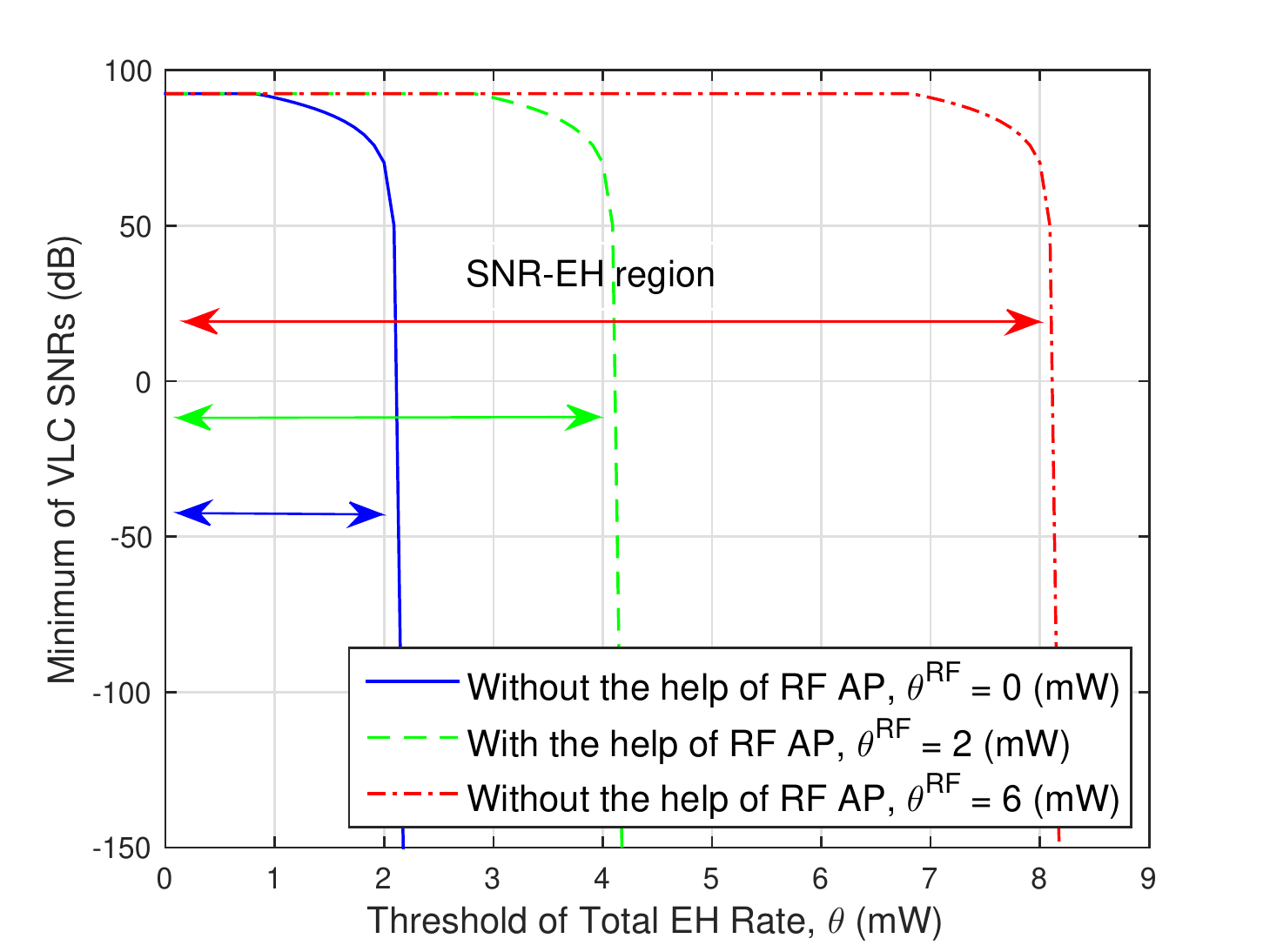}}
   \caption{
     Outperformance of the hybrid RF and lightwave power transfer network.
    }
    \label{fig:Support}
\end{figure}

In Fig. \ref{fig:Support}, the outperformance of the collaborative RF and lightwave resource allocation is shown. 
We explore the SNR-EH region of problem OP$_1$ over threshold $\theta$. Without the help of the RF AP, solving problem OP$_1$ becomes infeasible when $\theta$ is larger than 2.1 mW/s. This is because user 4 has the lowest SNR range and its maximum VLC EH is $2.1$ mW, as observed from Fig. \ref{fig:SNREH}.
One can see that the SNR-EH region is significantly enlarged when a higher level of $\theta^{RF}$ is given by the RF AP. In fact, the greater the amount of RF wireless power is, the more extensive the feasible region is. This implies a significant outperformance of the hybrid approach over the pure VLC one. 

\begin{figure}[t]
\centering
{\includegraphics[width=0.45\textwidth]{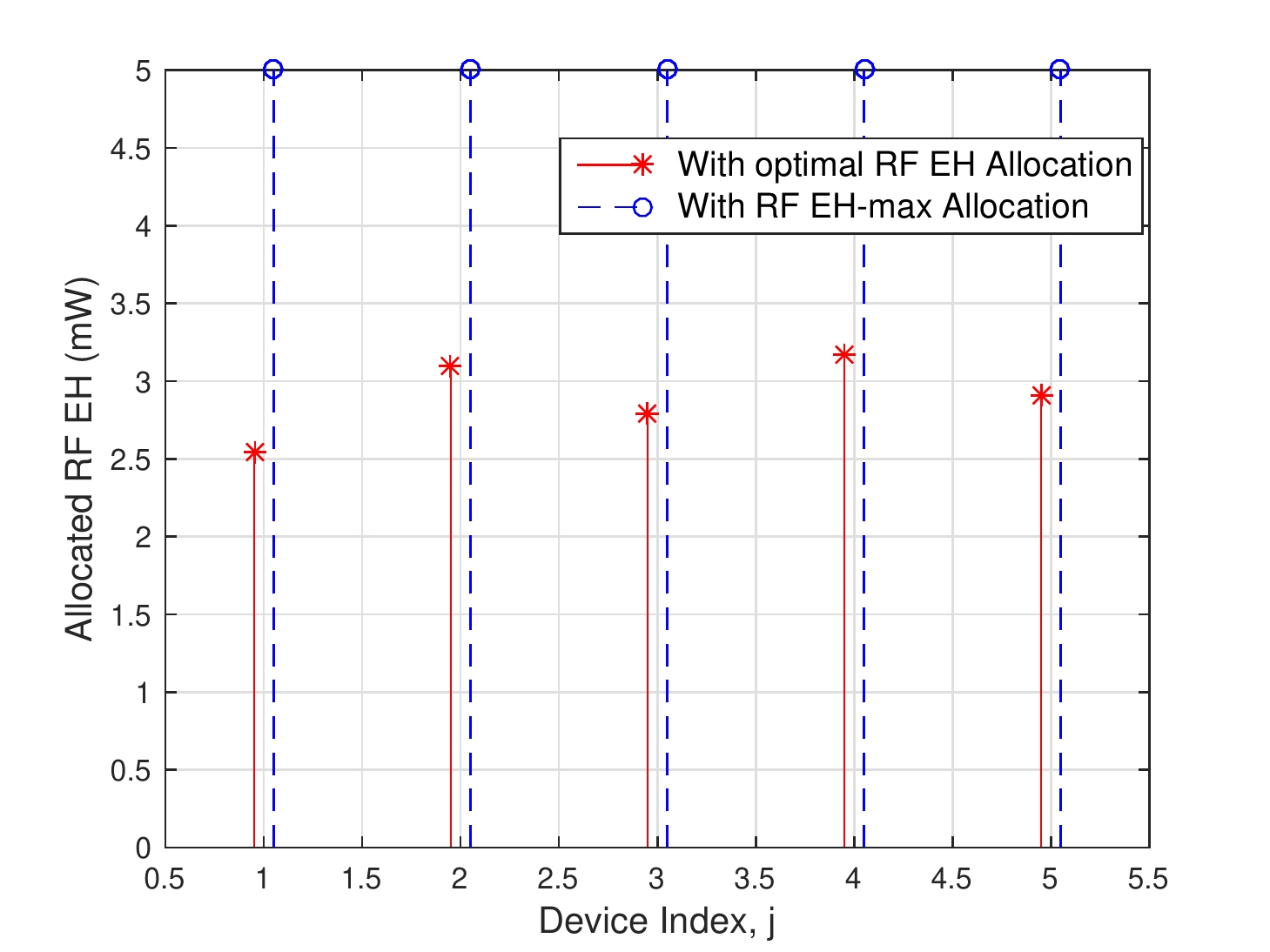}}
   \caption{
      Impact of EH rate allocation, $\theta^{RF} = 5$ and $\theta = 4$.
    }
    \label{fig:EHallocation}
\end{figure}

\begin{figure}[t]
\centering
{\includegraphics[width=0.45\textwidth]{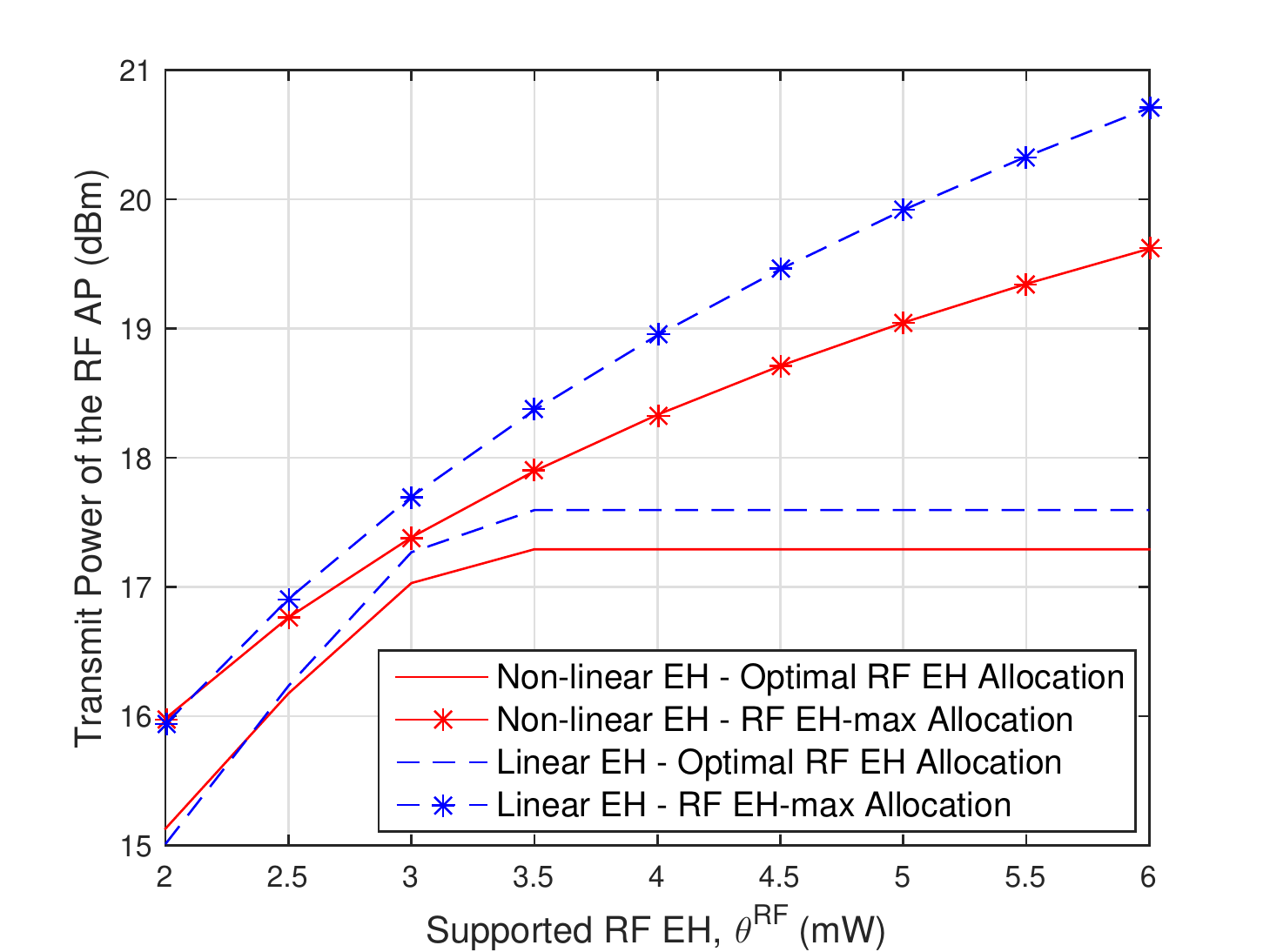}}
   \caption{
      Transmit power at the RF AP.
    }
    \label{fig:RFtransmitpower}
\end{figure}

In Fig. \ref{fig:EHallocation}, the importance of problem SubRF regarding the EH rate allocation is highlighted. Indeed, the manner of allocating all possible RF EH budget, termed as RH EH-max allocation, to each user is wasteful. Fig. \ref{fig:EHallocation} shows the optimal RH EH needed at each user to achieve the same minimum VLC SNR performance. In general, the optimal RF EH allocation can help avoid the transfer of redundant wireless power. It directly results in saving a significant amount of RF transmit power, which is clarified in the next simulation.

Fig. \ref{fig:RFtransmitpower} illustrates the amount of transmit power at the RF AP spent to supply the users with the preset EH threshold $\theta^{RF}$. In this simulation, the impact of the EH models and the RF EH allocation on transmit power is revealed. It is obvious that employing the optimal RH EH allocation leads to more power saving. Regarding the linear EH model, the energy conversion efficiency is set to 0.5. 
In this concern, it is observed that the power efficiency of the non-linear EH model is considerably higher than the idealistic linear model, which indicates the suitability of the proposed analysis for the applications of practical systems.

\begin{figure}[t]
\centering
{\includegraphics[width=0.45\textwidth]{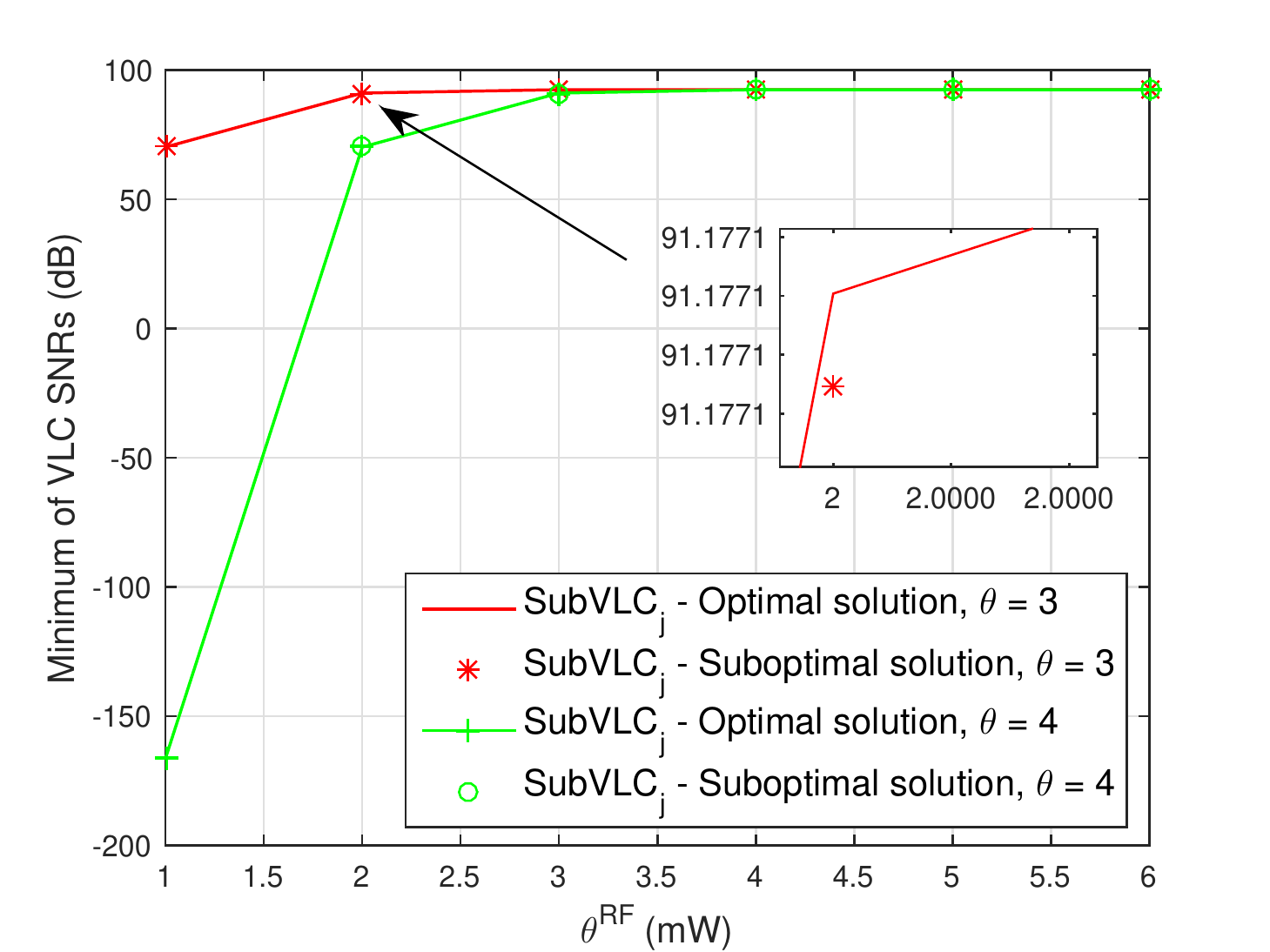}}
   \caption{
      Suboptimality of the solution given in Lemma 2 to problem \{SubVLC\}.
    }
    \label{fig:Suboptimal}
\end{figure}

Fig. \ref{fig:Suboptimal} presents the effectiveness of the suboptimal solution given in Lemma 2 of \{SubVLC$_j$\}. As observed, the suboptimal minimum of SNRs is slightly lower than the optimal one. In other words, the closed-form suboptimal solution can provide a very close system performance compared with the optimal solution. 
This can be explained by the fact that the approximation given in (50) is very tight since ${I_D} \ll {I_{j,G} (B)}$, i.e., $I_D = 10^{-9}$ while ${I_{j,G} (B)} \approx 10^{-3}$ . Thus, the optimality loss is negligible. Therefore, it is confirmed that the proposed suboptimal solution has not only low complexity but also high accuracy.

\begin{figure}[t]
\centering
{\includegraphics[width=0.45\textwidth]{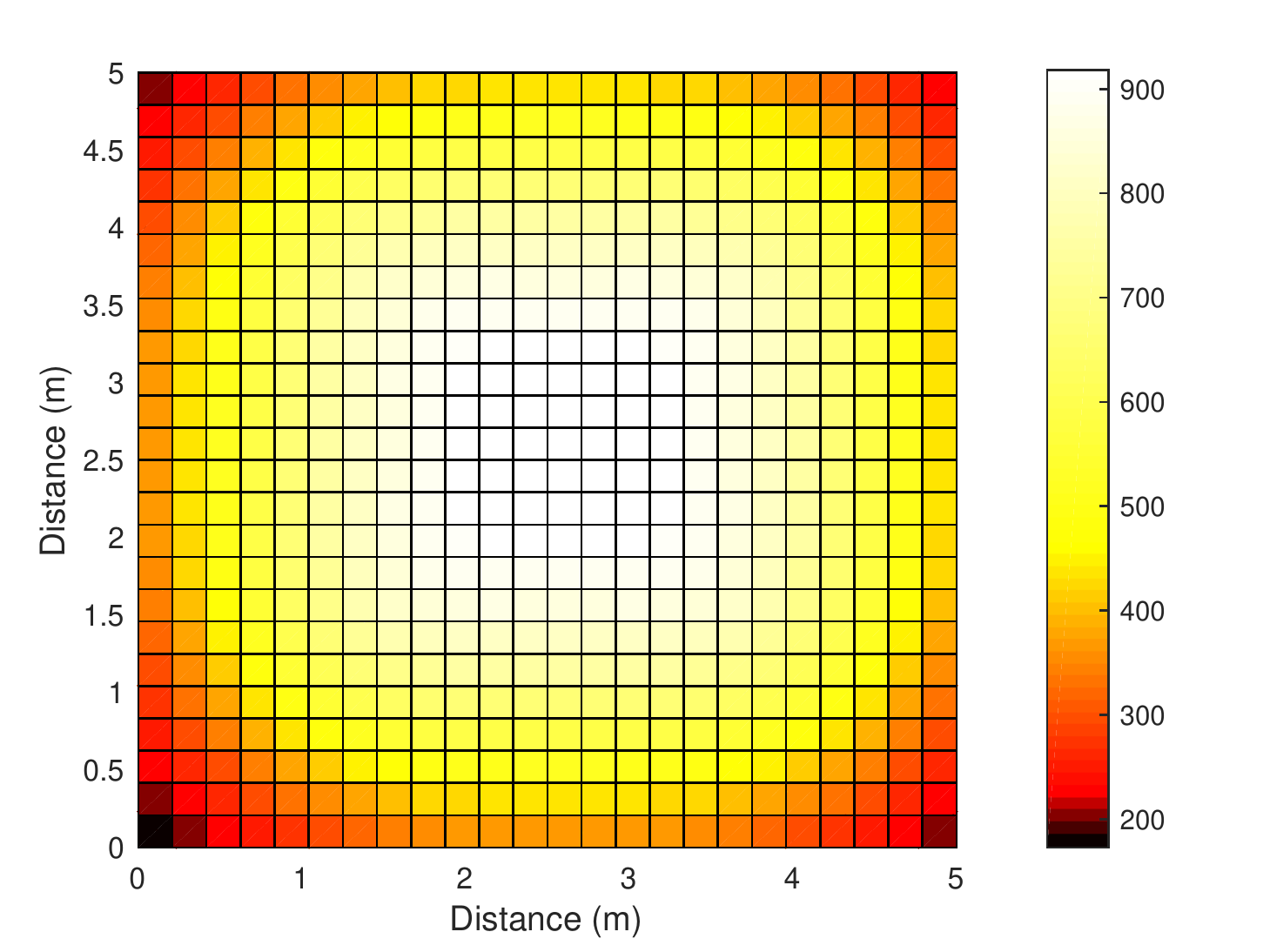}}
   \caption{
      Illuminance in the considered area.
    }
    \label{fig:Illuminance}
\end{figure}

In Fig. \ref{fig:Illuminance}, the illumination in the served area is shown to justify the practicality of the proposed system. Actually, the hybrid RF/VLC system should provide enough illumination for working or studying in the considered office environment.. In this regard, the illuminance can be calculated by a multiplication of the transmitted power and the typical efficacy of a LED. The transmitted power can be calculated by $N_{LED}*V_{LED}*B$ with $B = 0.0085$ A. According to \cite{Philips}, the typical efficacy is $90$ lm/W. By observing Fig. \ref{fig:Illuminance}, it can be seen that the achieved illuminance is approximately 920 lx at the center. Also, the illuminance in most of the area is more than 500 lx, which is the minimum value specified for office workers typing and reading documents by the European lighting standard \cite{Europeanlighting}.
Moreover, we can see that (i) different places in the room have different values of illumination, thus, some receivers cannot harvest sufficient energy, and (ii) the harvested energy by solely using VLC is limited.

\section{Conclusion and Future Potential Research}
In this work, we proposed a novel collaborative RF and lightwave resource allocation scheme for hybrid RF/VLC ultra-small cell networks. 
In the proposed scheme, the optical transmitters play the main role and deliver both the lightwave information and energy signals whereas the RF AP is an assistant and transfers wireless power over RF signals.
Thus, the minimum of the VLC SNRs is maximized under the constraints of the sum of light and RF EH performances.
On this basis, we provide the algorithms to solve the problems associated with the group of the optical transmitters and the RF AP optimally. Further, we derive a simple closed-form suboptimal solution to the optical transmitters' problem to facilitate the solving process. Particularly, the suboptimal solution can be seen as an efficient low-complexity alternative with high accuracy when the network needs to update solutions instantly.
The numerical results indicate that the collaborative scheme significantly improves the overall network performance in terms of VLC SNR and EH metrics while maintaining an appropriate illumination in the area.

For future research, an extended scenario, in which multiple RF AFs cooperate with multiple optical transmitters, should be considered to improve both ID and EH performances and reduce the transmit power. However, this scenario might lead to the increase in the hardness of finding optimal and suboptimal resource allocations.
Furthermore, the system model in this work can also be developed with invisible light as an additional energy and information source, such as infrared light. Managing the invisible light band might be more flexbile than the visible one since it does not affect illumination. Nevertheless, resource allocation schemes need to carefully take human eye safety regulations into account.

\appendices

\section{Proof of Proposition 1}
For convenience, we use $\mathtt {SNR}^{\mathtt{s}}_{oi,j}(B)$ to denote that the VLC SNR at any user is a function of $B$. Our aim is to maximize the minimum of the VLC SNRs, i.e., $\max \hspace{2pt} \min \hspace{2pt}  \mathtt {SNR}^{\mathtt{s}}_{oi,j}(B)$.

Since $\mathtt {SNR}^{\mathtt{s}}_{oi,j}(B)$ is an increasing function, inreasing $B$ leads to increasing the VLC SNRs at all the users. Thus, maximizing the minimum of VLC SNRs is equivalent to simply maximizing the VLC SNR at the user with the worst VLC SNR, mathematically represented as 
\begin{align}
\max \hspace{2pt} \min \hspace{2pt}  \mathtt {SNR}^{\mathtt{s}}_{oi,j}(B) = \max \hspace{2pt}  \mathtt {SNR}^{\mathtt{s}}_{oi,{\bar j}}(B),
\end{align}
where $\bar j$ denotes the user with the worst VLC SNR.

Furthemore, according to \eqref{eq:maxVLCSINRb}, there is a relationship between the VLC SNR and the VLC EH. It can be seen that maximizing the VLC SNR implies minimizing the VLC EH. Hence, we have 
\begin{align}
\max \hspace{2pt}  \mathtt {SNR}^{\mathtt{s}}_{oi,{\bar j}}(B) = \min \hspace{2pt} \mathtt {EH}_{\bar j}^{VLC}(B).
\end{align}

Next, due to constraint \eqref{eq:maxVLCSINRb}, reducing $\mathtt {EH}_{\bar j}^{VLC}(B)$ is achieved by adding more $\mathtt {EH}_{\bar j}^{RF}$. Generally speaking, increasing the RF EH reduces the contribution of the VLC EH, and increases the achievable amount of the VLC SNRs. Therefore, the RF EH allocated to the user with the worst VLC SNR determines the value of the minimum of the VLC SNRs, and therefore determines $B$ for all the users. This completes the proof.

\section{Proof of Lemma 1}
To demonstrate the rank-one issue, we can have some following expressions based on the Karush-Kuhn-Tucker (KKT) condition as \cite{StephenBoyd2004}
\begin{align}
{\mathbf I} - \gamma_j {\mathbf G}_{j} - {\pmb \Xi}_j &= {\mathbf 0}, \label{eq:KKTa} \\
{\mathbf W}_j {\pmb \Xi}_j &= {\mathbf 0}, \label{eq:KKTb}
\end{align}
where $\gamma_j$ and ${\pmb \Xi}_j $ are dual variables associated with constraints \eqref{eq:SDRproba} and \eqref{eq:SDRprobb}, respectively. 

On one hand, according to \eqref{eq:SDRproba}, \eqref{eq:SDRprobb}, and \eqref{eq:KKTb}, it is implied that ${\mathbf W}_j \neq {\mathbf 0}$. Thus, we infer that 
\begin{align}\label{eq:cond1}
\text{rank}({\pmb \Xi}_j ) \le M_T - 1.
\end{align}

On the other hand, based on \eqref{eq:KKTa}, we have
\begin{align}\label{eq:cond2}
\text{rank}({\pmb \Xi}_j ) &= \text{rank}({\mathbf I} - \gamma_j {\mathbf G}_{j}) \ge  M_T - 1.
\end{align}

Taking \eqref{eq:cond1} and \eqref{eq:cond2} into account, it is clear that
\begin{align}
\text{rank}({\pmb \Xi}_j ) = M_T - 1.
\end{align}
Next, in light of \eqref{eq:KKTb}, the rank of ${\mathbf W}_j$ can be calculated by
\begin{align}
\text{rank}({\mathbf W}_j  ) \le M_T - \text{rank}({\pmb \Xi}_j ).
\end{align}

Since $\text{rank}({\mathbf W}_j  ) \neq 0$ due to ${\mathbf W}_j \neq {\mathbf 0}$, we conclude that
\begin{align}
\text{rank}({\mathbf W}_j  ) = 1.
\end{align}
This completes our proof.

\section{Proof of Lemma 2}
The key to achieving a closed-form solution for problem \eqref{eq:VLCsubProblem2} relies on dealing with constraint \eqref{eq:VLCsubProblemb1}. We recall the constraint as follows
\begin{align}\label{eq:lemma31}
f V_0 I_{{\bar j},G} (B)  \text{ln} \left(1 + \dfrac{I_{{\bar j},G} (B)}{I_D}\right) \ge {\theta -  \mathtt {EH}_{\bar j}^{\star RF}}{}.
\end{align}
Initially, we aim to transform the constraint into a simplified form of $B$ greater than a constant.
Unfortunately, this is difficult to achieve. Then, deriving a closed-form optimal solution is intractable. However, a closed-form suboptimal solution can be obtained by relaxing constraint \eqref{eq:VLCsubProblemb1}. Our method is presented as follows.

Since ${I_D} \ll {I_{{\bar j},G} (B)}$ in practice, we can have an approximation as
\begin{align}\label{eq:lemma32}
\text{ln} \left(1 + \dfrac{I_{{\bar j},G} (B)}{I_D}\right) \approx \text{ln} \left(\dfrac{I_{{\bar j},G} (B)}{I_D}\right).
\end{align}
Thus, \eqref{eq:lemma31} becomes
\begin{align}
I_{{\bar j},G} (B)  \text{ln} \left( \dfrac{I_{{\bar j},G} (B)}{I_D}\right) \ge \frac{\theta-  \mathtt {EH}_{\bar j}^{\star RF}}{f V_0 },
\end{align}

After some manipulations, $I_{{\bar j},G} (B)$ can be computed as
\begin{align}
I_{{\bar j},G} (B)  \ge \frac{\theta -  \mathtt {EH}_{\bar j}^{\star RF}}{f V_0 \mathcal W \left( \frac{\theta-  \mathtt {EH}_{\bar j}^{\star RF}}{f V_0 I_D}\right)  } 
\end{align}
where $\mathcal W (.)$ is the Lambert function. Accordingly, we have
\begin{align}
B \ge \frac{\theta -  \mathtt {EH}_{\bar j}^{\star RF}}{3\nu N_{LED}V_{LED}f V_0 \mathcal W \left( \frac{\theta -  \mathtt {EH}_{\bar j}^{\star RF}}{f V_0 I_D}\right) \sum\limits_{o}^{\mathcal O} \sum\limits_{i}^{M_I} h_{oi,{\bar j}}}= \Upsilon.
\end{align}

Accordingly, problem \eqref{eq:VLCsubProblem2} becomes
\begin{subequations}\label{eq:VLCsubProblem3}
\begin{align}
	  \underset{  B}  \min \quad &  B  \quad  \label{eq:VLCsubProblem3a} 
\\
\text{s.t.:}  	
			\quad & B \ge \Upsilon,  \label{eq:VLCsubProblem3c}	\\			
			\quad &  \dfrac{I_H + I_L}{2} \le B \le I_H.
\end{align}
\end{subequations}

As observed, the closed-form suboptimal solution can be computed as follows
\begin{align}
B^{\times} &= \min \left\{ \max \{ \Upsilon, I_L\},  I_H \right\}, \\
A^{\times \mathtt{s}}_{oi,j} &= B - B^{\times }. \quad ( \forall j)
\end{align}
Thus, our proof is completed.

\bibliographystyle{IEEEtran}
\bibliography{IEEEabrv,REF}

\end{document}